\documentclass[fleqn,10pt]{wlscirep}
\usepackage{amssymb}
\usepackage{amsmath}
\usepackage{graphicx}
\usepackage{booktabs}
\usepackage{multirow}
\usepackage{float}
\usepackage{algorithm}
\usepackage{algorithmicx}
\usepackage{algpseudocode}
\usepackage{bm}
\usepackage{longtable}
\usepackage{hyperref}
\usepackage{placeins}
\usepackage{booktabs}
\usepackage{multirow}
\usepackage{tabularx}
\usepackage{subcaption}

\title{Refracted Light Interaction in Turbulent Bubbling Water (RLITBW): A Macroscopic Fluid--Optic Entropy Source}
\author[1,*]{Nirjhar Debnath}
\author[1]{Dwaipayan Datta}
\author[1]{Kousik Dasgupta}

\affil[1]{Department of Computer Science and Engineering,
Kalyani Government Engineering College, Kalyani, 741253, India}

\affil[*]{nirjhardebnath2006@gmail.com}

\begin{abstract}
The demand for high-quality, unpredictable random numbers is a fundamental requirement in cryptography, stochastic simulation, and optimization. While pseudo-random number generators (PRNGs) are computationally efficient, their deterministic nature limits their suitability for security-critical applications. True random number generators (TRNGs), although physically grounded, often rely on expensive quantum or tightly controlled electronic phenomena. This paper introduces a low-cost, macroscopic TRNG based on \emph{Refracted Light Interaction in Turbulent Bubbling Water (RLITBW)}. The proposed system exploits compound classical chaos arising from multiphase fluid dynamics and time-varying optical refraction. A physical–mathematical model is developed to describe the cascade of non-linear processes—from stochastic bubble nucleation and turbulent ascent to chaotic optical path scrambling—that collectively amplify microscopic uncertainties into measurable entropy.The raw optical signal is digitized and processed using a provably secure entropy-conditioning pipeline based on Toeplitz universal hashing, followed by deterministic cryptographic expansion. The chaotic nature of the physical source is empirically validated using phase-space reconstruction, Lyapunov exponent estimation, autocorrelation analysis, and entropy metrics. The conditioned output successfully passes the full NIST SP 800-22 statistical test suite and nonlinear dynamical measures including Lyapunov exponents and sample entropy. Beyond statistical validation, the generated randomness is applied to population-based optimization algorithms, demonstrating practical usability as a replacement for conventional PRNGs. Finally, deployment architectures and scalability considerations are discussed, positioning RLITBW as an accessible, reproducible, and economically viable entropy source for real-world systems.

\end{abstract}

\keywords{Fluid optics \sep Optical entropy \sep Turbulent Liquid \sep Randomness in Light direction \sep True Random Number Generator}

\begin{document}
\flushbottom
\maketitle

\section*{Introduction}
Randomness plays an important role in many areas of modern science and security. From generating cryptographic keys and initializing secure protocols to performing Monte Carlo simulations and stochastic modeling, the availability of high-quality, unpredictable number sequences is critical \cite{b1}. Computational algorithms for this purpose are broadly classified into two families: Pseudo-Random Number Generators (PRNGs) and True Random Number Generators (TRNGs).

PRNGs are deterministic algorithms that, given an initial seed, produce long sequences of numbers that exhibit statistical properties of randomness. While computationally efficient and reproducible, their deterministic nature is a critical vulnerability; if the seed and algorithm are known, the entire sequence is predictable \cite{b2}. This makes PRNGs unsuitable for applications where unpredictability is paramount.

TRNGs, or physical random number generators, derive their output from an inherently unpredictable physical process, often termed an entropy source \cite{b3}. The unpredictability of these generators is rooted in the laws of physics, making their output theoretically impossible to reproduce. Common entropy sources include thermal noise in resistors, the timing of radioactive decay, and clock jitter in electronic circuits \cite{b4}. The current gold standard is represented by Quantum Random Number Generators (QRNGs), which leverage the fundamental indeterminacy of quantum phenomena, such as photon splitting or quantum tunneling, to produce provably unpredictable sequences \cite{b5}.

While effective, these physical generators present a trade-off. Electronic noise-based TRNGs can be sensitive to environmental conditions or adversarial manipulation, requiring careful post-processing. QRNGs, though theoretically ideal, demand specialized and expensive hardware, such as single-photon detectors, placing them out of reach for many low-cost or embedded applications. This reveals a gap: a need for TRNGs that are physically robust and truly unpredictable, yet also low-cost, accessible, and easily constructed from common components.

This paper proposes a novel TRNG system that fills this gap by harnessing the complex dynamics of \textbf{macroscopic classical chaos}. The proposed system, Refracted Light Interaction in Turbulent Bubbling Water (RLITBW), uses the chaotic interplay of multiphase fluid dynamics and optical refraction as its entropy source. This work posits that such a system, built from inexpensive, affordable components, can function as a high-entropy source.

The primary contributions of this work are as follows:
\begin{itemize}
    \item The design and physical–mathematical modeling of a novel macroscopic entropy source based on fluid–optic chaos.
    \item Experimental realization of the RLITBW system using low-cost, commonly available components.
    \item Empirical validation of chaotic behavior through time-series analysis, phase-space reconstruction, Lyapunov exponent estimation, and entropy measures.
    \item A standards-aligned entropy conditioning pipeline using universal hashing and deterministic cryptographic expansion.
    \item Statistical verification of the generated bitstreams using the NIST SP 800-22 test suite.
    \item Demonstration of practical applicability through integration with representative optimization algorithms and proposal of scalable deployment architectures.
\end{itemize}

The remainder of this paper is organized as follows. The \textit{Literature Review} explores existing TRNG technologies and identifies the research gap addressed in this work. The section on \textit{RLITBW TRNG} presents the physical--mathematical and algorithmic model of the proposed entropy source. The \textit{Experimental Setup and Measurement Protocol} describes the experimental design, data acquisition, and processing pipeline. The section on \textit{Entropy Extraction, Conditioning, and Bitstream Expansion} details the methodology for generating high-quality random bitstreams. The \textit{Statistical Evaluation} presents the results of randomness testing and analysis. The section on \textit{Applications to Optimization Algorithms} demonstrates the practical use of the generated randomness. \textit{Practical Deployment and Scalability} discusses implementation aspects and scalability considerations, while \textit{Discussion and Performance Analysis} evaluates performance, cost, and robustness. Finally, the \textit{Conclusion} summarizes the findings and outlines directions for future work.

\section*{Literature Review: The Landscape of Entropy Sources}

The quality and reliability of a true random number generator (TRNG) are fundamentally determined by its underlying physical entropy source. Over the past several decades, research in this domain has largely converged around three principal classes of entropy-generating phenomena: electronic and thermal noise, quantum-mechanical processes, and deterministic chaotic systems. Each category offers distinct advantages and limitations in terms of unpredictability, implementation complexity, scalability, and cost.

\subsection*{Electronic and Thermal Noise}

The earliest and most widely deployed TRNGs exploit stochastic noise processes inherent to electronic components. These include thermal (Johnson--Nyquist) noise in resistors, shot noise in semiconductor junctions, and avalanche noise in Zener diodes \cite{b3}. In modern digital systems, particularly field-programmable gate arrays (FPGAs) and application-specific integrated circuits (ASICs), entropy is commonly harvested from timing jitter in free-running ring oscillators (ROs) \cite{b7}. In RO-based TRNGs, microscopic perturbations arising from thermal noise and power supply fluctuations induce small variations in oscillation periods, which are sampled to generate random bitstreams.

While such approaches are attractive due to their ease of integration and low hardware overhead, they are not without limitations. Electronic noise sources can exhibit bias, temporal correlation, and sensitivity to environmental conditions such as temperature and supply voltage. Moreover, several studies have demonstrated the susceptibility of purely electronic TRNGs to fault injection and active manipulation attacks, necessitating careful design and continuous health monitoring.

\subsection*{Quantum Random Number Generators (QRNGs)}

Quantum random number generators represent the current state-of-the-art in randomness generation, leveraging the fundamental probabilistic nature of quantum measurement outcomes \cite{b5}. In contrast to classical systems, the unpredictability of a quantum process persists even under complete knowledge of the system state, offering the strongest theoretical guarantees of randomness. Typical QRNG implementations are optical, exploiting phenomena such as photon path splitting at beam splitters, photon arrival-time measurements, phase noise in lasers, or quantum tunneling effects \cite{b5,b8}.

Despite their conceptual elegance and strong security guarantees, QRNGs present practical challenges. They require specialized and often delicate hardware components, including single-photon detectors and stabilized laser sources, which increase cost, power consumption, and system complexity. These constraints can limit their adoption in resource-constrained or large-scale deployments, motivating continued exploration of alternative physical entropy sources.

\subsection*{Chaotic Systems}

A third major class of TRNGs is based on deterministic chaotic systems. Chaotic systems are nonlinear dynamical systems characterized by extreme sensitivity to initial conditions, commonly referred to as the ``butterfly effect'' \cite{b9}. Although deterministic in principle, their long-term evolution becomes practically unpredictable due to unavoidable uncertainties in initial state measurement and environmental perturbations. As a result, chaos-based systems offer a compelling intermediate position between purely stochastic noise sources and fundamentally probabilistic quantum processes.

Recent experimental advances further highlight the diversity of chaos-based entropy sources, including novel physical platforms such as spin-crossover self-oscillating systems, which demonstrate emerging hardware modalities for TRNG implementation \cite{spin_crossover2024}. Existing research on chaos-based TRNGs has primarily focused on two broad categories:

\begin{itemize}
	\item \textbf{Chaotic Lasers:} Semiconductor lasers subjected to delayed optical feedback can be driven into chaotic regimes, producing broadband, aperiodic intensity fluctuations. These optical chaos signals can be digitized at very high sampling rates, enabling random bit generation at gigabit-per-second scales \cite{b10}. More recent optical physical unclonable function (PUF) and speckle-based systems further demonstrate that macroscopic optical complexity can serve as a reliable entropy source for high-throughput random number generation \cite{opt_puf2021}. In particular, speckle pattern randomness generated in multimode optical systems has been shown to satisfy cryptographic randomness requirements after appropriate post-processing \cite{speckle_iet2022,optical_speckle_trng2020}.
	
	\item \textbf{Chaotic Circuits:} Electronic circuits implementing nonlinear dynamical equations—such as Chua's circuit and related chaotic oscillators—constitute another well-studied entropy source. The resulting non-periodic voltage or current signals are sampled and digitized to produce random bitstreams \cite{b9,b11}. While effective, these systems remain confined to the electronic domain and are subject to similar environmental sensitivities as conventional noise-based TRNGs.
\end{itemize}

\subsection*{Macroscopic Optical and Sensor-Based Entropy Sources}

Beyond microscale electronic and quantum implementations, recent work has increasingly explored macroscopic physical systems as viable entropy sources, particularly those involving optical and fluid-dynamic processes. Optical speckle–based random number generators exploit the extreme sensitivity of interference patterns to microscopic perturbations, and have been shown to produce statistically robust random sequences after suitable conditioning \cite{opt_speckle2021}. Related approaches leverage optical turbulence and dynamic refractive index fluctuations to amplify small-scale randomness into measurable macroscopic intensity variations suitable for entropy extraction \cite{opt_turbulence2020}. Collectively, these studies suggest that complex light propagation through dynamically evolving media can serve as a practical source of physical randomness without reliance on quantum-scale effects.

In parallel, turbulent fluid systems themselves have been examined from an information-theoretic perspective, revealing sustained entropy generation driven by nonlinear interactions and sensitivity to initial conditions \cite{fluid_entropy2022}. When combined with optical sensing, such fluid-dynamic randomness can be transduced into electrical signals whose unpredictability is further shaped by sensor noise and mixed-signal imperfections. Recent investigations demonstrate that noise processes intrinsic to photodetectors and analog-to-digital converters can contribute meaningful entropy when appropriately modeled and conservatively estimated \cite{sensor_noise2020,entropy_estimation2021}.

At the system level, growing emphasis has been placed on robust validation methodologies for physical random number generators, including continuous health testing and runtime entropy monitoring. Surveys of modern TRNG architectures highlight the importance of combining physical entropy sources with online health tests to detect degradation, bias, or environmental drift during long-term operation \cite{health_tests2023}. From a dynamical systems perspective, the interaction between deterministic nonlinear dynamics and stochastic perturbations—often described as stochastic chaos—has been identified as a key mechanism by which macroscopic systems can exhibit effectively unpredictable behavior \cite{chaos_noise2020}. These developments collectively motivate the exploration of optically sensed fluid-dynamic systems as scalable macroscopic entropy sources, closely aligning with the refracted-light–based turbulence mechanism proposed in this work \cite{macro_trng2024}.

\subsection*{Lava Lamps as a Source of Randomness}

A prominent real-world demonstration of macroscopic physical randomness is Cloudflare's ``LavaRand'' system, in which a wall of lava lamps serves as an entropy source for cryptographic seeding \cite{lava_rand_patent,cloudflare_lavarand}. In this setup, cameras continuously capture images of the evolving wax blobs within the lamps. The wax motion is driven by heat transfer, convection, and fluid dynamics, producing complex shapes and trajectories that are effectively impossible to predict or reproduce.

The captured images contain substantial natural randomness in their pixel intensities and spatial patterns. Cloudflare combines this image-derived entropy with other hardware sources before applying cryptographic randomness extractors to generate uniform random seeds. LavaRand is significant because it demonstrates that visible, macroscopic chaotic processes can be harnessed as reliable TRNG sources, without reliance on specialized quantum hardware. This approach closely parallels the philosophy underlying the RLITBW system, wherein chaotic fluid motion and optical distortion are similarly exploited to generate physical entropy.

\subsection*{Identified Research Gap and Contribution}

The existing literature is heavily weighted toward either quantum-mechanical phenomena or microscale electronic chaos. While these approaches are well established, they typically require specialized hardware or remain confined to electronic implementations. In contrast, the potential of macroscopic, fluid-dynamic, and optically mediated chaotic systems as primary entropy sources remains comparatively underexplored.

This work directly addresses this gap by proposing the RLITBW system as a new class of chaos-based TRNG characterized by:
\begin{enumerate}
	\item \textbf{Macroscopic Accessibility:} The entropy-generating phenomenon is easily observable and constructed from low-cost, commodity components, including an air pump, a light source, a photometric sensor, and a simple fluid vessel.
	\item \textbf{Physical Grounding:} Randomness arises from well-understood nonlinear governing principles of fluid dynamics and optics, rather than opaque electronic jitter alone.
	\item \textbf{Compound Chaos:} The system combines turbulent multiphase fluid dynamics with dynamic optical path scrambling caused by light refraction through a time-varying, non-homogeneous medium.
\end{enumerate}

By investigating the feasibility of harnessing this accessible yet complex physical process, this work aims to demonstrate that statistically robust true random numbers can be generated from macroscopic fluid–optical chaos, offering a practical alternative to existing TRNG paradigms.

%
%
%

\FloatBarrier
\section*{The RLITBW TRNG: A Physical–Mathematical and Algorithmic Model}

The RLITBW system derives unpredictability from a cascade of interacting chaotic
processes. Although each physical process can be individually described by known
equations, their coupling produces a high-dimensional system that is practically
impossible to predict. This section introduces the full model and provides 
clear definitions of all variables.

\subsection*{Variable and Symbol Definitions}

For clarity, Table~\ref{tab:notation_rlitbw} summarizes all variables and
mathematical symbols used in the RLITBW physical--mathematical model.
Each variable is referenced in the models that follow.

\begin{table*}[!t]
\centering
\small
\setlength{\tabcolsep}{6pt}
\renewcommand{\arraystretch}{1.15}

\begin{tabular}{ll|ll}
\toprule
\textbf{Symbol} & \textbf{Description} &
\textbf{Symbol} & \textbf{Description} \\
\midrule

$\rho_w$ & Density of water
& $n_w$ & Refractive index of water \\

$\rho_a$ & Density of air
& $n_a$ & Refractive index of air \\

$\rho$ & Fluid density in Navier--Stokes equations
& $\theta_w$ & Refraction angle in water \\

$g$ & Acceleration due to gravity
& $\theta_a$ & Refraction angle in air \\

$\sigma$ & Surface tension coefficient
& $\Psi(\mathbf{u},t)$ & Optical intensity at sensor plane \\

$d$ & Diameter of air orifice
& $\Psi_0(\mathbf{u}_0)$ & Incident irradiance \\

$P_{\text{air}}(t)$ & Air pressure input
& $\mathbf{u}$ & Sensor-plane coordinate \\

$\delta P(t)$ & Pressure fluctuations
& $\mathbf{u}_0$ & Input-plane coordinate \\

$V_i(t)$ & Bubble volume
& $\mathcal{T}_t$ & Time-dependent ray mapping \\

$t_i$ & Bubble detachment time
& $D\mathcal{T}_t$ & Jacobian of ray mapping \\

$\Delta t_i$ & Inter-bubble interval
& $A$ & Sensor active area \\

$\mathbf{r}_i(t)$ & Bubble position vector
& $\Delta t$ & Sensor integration time \\

$\mathbf{v}_i(t)$ & Bubble velocity
& $\tau$ & Integration variable \\

$\mathbf{v}$ & Fluid velocity field
& $I(t)$ & Integrated sensor intensity \\

$\mathbf{X}_i(t)$ & Bubble state vector
& $\kappa$ & Bubble growth coefficient \\

$\mathbf{S}(t)$ & Global system state
& $k$ & Ray index \\

$\mathbf{S}'(t)$ & Perturbed state
& $X_t$ & Digitized sensor output \\

$\delta_0$ & Initial perturbation
& $B(t)$ & Extracted bitstream \\

$\nu$ & Kinematic viscosity
& $\nabla$ & Gradient operator \\

$p$ & Fluid pressure
& $\nabla^2$ & Laplacian operator \\

$\mathbf{f}_{\text{buoy}}$ & Buoyancy force density
& $\|\cdot\|$ & Euclidean norm \\

$\mathbf{f}_{\text{bubble}}$ & Bubble interaction force
& $\det(\cdot)$ & Matrix determinant \\

$F_b(t)$ & Total buoyancy force
& $\mathrm{ADC}(\cdot)$ & Analog-to-digital conversion \\

$F_s$ & Surface tension force
& $\mathrm{ExtractBits}(\cdot)$ & Bit extraction function \\

$\lambda$ & Lyapunov exponent
& $\mathbb{P}(t)$ & 3D bubble distribution \\

\bottomrule
\end{tabular}

\caption{Variables, symbols, and mathematical notation used in the RLITBW physical--mathematical model.}
\label{tab:notation_rlitbw}
\end{table*}

\subsection*{Stage 1: Bubble Nucleation as a Stochastic Trigger Process}

A bubble begins forming at the orifice as air inflates a spherical cap.
Its volume evolves according to:

\begin{equation}
    \frac{dV_i}{dt}
    = \kappa \big( P_{\text{air}}(t) + \delta P(t) \big),
\end{equation}

where $P_{\text{air}}(t)$ is nominal pump pressure, $\delta P(t)$ is small random mechanical/vibrational noise, and $\kappa$ is a proportionality constant representing a simplified linear growth model that captures the qualitative dependence of bubble inflation rate on pressure differential.

A bubble detaches when buoyancy exceeds surface tension:

\begin{equation}
    F_b(t) = (\rho_w - \rho_a) g V_i(t), \qquad
    F_s = \pi d \sigma.
\end{equation}

Detachment rule:

\begin{equation}
    F_b(t_i) > F_s.
\end{equation}

This produces a random timing sequence $\Delta t_i$ and random initial bubble sizes.
These variations serve as the \textbf{seed entropy} $\delta_0$.

\subsection*{Stage 2: Chaotic Amplification via Multiphase Turbulent Flow}

Once released, the bubbles enter a turbulent water column.
The fluid motion is governed by the incompressible Navier–Stokes equations \cite{batchelor2000}:

\begin{equation}
\frac{\partial \mathbf{v}}{\partial t}
+ (\mathbf{v} \cdot \nabla) \mathbf{v}
= -\frac{1}{\rho} \nabla p
+ \nu \nabla^2 \mathbf{v}
+ \mathbf{f}_{\text{buoy}}
+ \mathbf{f}_{\text{bubble}}.
\end{equation}

Here:

\begin{itemize}
    \item $(\mathbf{v} \cdot \nabla) \mathbf{v}$ is a nonlinear convective term creating turbulence,
    \item $\mathbf{f}_{\text{bubble}}$ includes wake interactions, collisions, and drag,
    \item the system has many interacting bodies (bubbles),
    \item the Reynolds number \cite{reynolds1883}, \cite{white2011} is high, ensuring turbulence.
\end{itemize}

\textbf{Note:} This formulation represents single-phase flow with volumetric body forces to model bubble-induced effects. A complete two-phase computational fluid dynamics (CFD) treatment would be more rigorous but is beyond the scope of this entropy-source characterization.

Because the system has a positive Lyapunov exponent $\lambda$ \cite{wolf1985}:

\begin{equation}
    \|\mathbf{S}(t) - \mathbf{S}'(t)\|
    \approx \delta_0 e^{\lambda t},
\end{equation}

microscopic differences in Stage 1 grow exponentially.
Thus the bubble positions, velocities, and shapes become unpredictable.

\subsection*{Stage 3: Optical Chaos via Time-Varying Refractive Micro-Lensing}

As bubbles rise and continuously deform, the water--air interfaces behave as a time-varying ensemble of refractive surfaces that repeatedly bend incident light rays.
Whenever a ray crosses a water--air boundary, its direction changes according to Snell's law (angles measured with respect to the local surface normal) \cite{hecht2017,bornwolf}:
\begin{equation}
n_w \sin\theta_w = n_a \sin\theta_a .
\end{equation}
As the interface geometry and local normals evolve in time, two rays that are initially close can undergo different sequences of refractions, yielding strong spatio-temporal modulation of the transmitted light.

For a simple and physically correct model, we treat the bubbling column as a \emph{piecewise-homogeneous} optical medium with refractive index
$n(\mathbf{x},t)\in\{n_w,n_a\}$, where $n_w$ applies in water and $n_a$ inside bubbles.
Within each homogeneous region (constant $n$), rays propagate approximately along straight lines; when a ray intersects a moving water--air interface, its direction is updated by Snell's law \cite{bornwolf}.
The cumulative effect of these time-dependent refractions defines a time-dependent ray mapping
\(\mathcal{T}_t\) from an incident ray coordinate \(\mathbf{u}_0\) (at an input plane) to the intersection point \(\mathbf{u}\) on the sensor plane.

\textbf{Geometric optics approximation:} This treatment is valid when the characteristic bubble diameter is much larger than the wavelength of visible light, which is satisfied in our experimental setup where typical bubble sizes are on the order of millimeters.

In geometric optics, the sensor-plane irradiance can be expressed via conservation of ray (energy) density: intensity increases where rays locally converge and decreases where they diverge \cite{stavroudis1972,berryupstill1980}.
Accordingly, if multiple rays indexed by $k$ reach the same sensor location $\mathbf{u}$, the irradiance is given by the standard Jacobian (ray-density) relation \cite{stavroudis1972}:
\begin{multline}
\Psi(\mathbf{u},t)=\sum_{k}\Psi_0(\mathbf{u}_{0,k})
\Big|\det\!\big(D\mathcal{T}_t(\mathbf{u}_{0,k})\big)\Big|^{-1}, \\
\mathbf{u}=\mathcal{T}_t(\mathbf{u}_{0,k}).
\end{multline}
where $\Psi_0$ is the incident irradiance distribution and $D\mathcal{T}_t$ is the Jacobian of the mapping.
As bubble motion and deformation cause rapid fluctuations in $\mathcal{T}_t$, the resulting $\Psi(\mathbf{u},t)$ exhibits transient focusing/defocusing and caustic-like bright/dark structures \cite{berryupstill1980}.
This time-varying optical field is then integrated by the photometric sensor in Stage~4.

\subsection*{Stage 4: Sensor Integration and Scalar Quantization}

The sensor integrates all incoming light across its area $A$ and over its
exposure window $\Delta t$:

\begin{equation}
    I(t)
    =
    \int_A
    \int_t^{t+\Delta t}
        \Psi(\mathbf{u}, \tau)
    \, d\tau\, dA.
\end{equation}

This integration collapses the entire chaotic optical field into one scalar number.
The sensor then converts $I(t)$ into a 16-bit value:

\begin{equation}
    X_t = \text{ADC}(I(t)).
\end{equation}

Here sensor noise and ADC quantization effects also contribute to the randomness. Finally, raw bits are extracted:

\begin{equation}
    B(t) = \text{ExtractBits}(X_t),
\end{equation}

where typical extraction uses lower bits, which carry more entropy.

\subsection*{Algorithmic Summary}

\begin{algorithm}[ht]
\caption{Physical-to-Digital Entropy Pipeline}
\begin{algorithmic}[1]
\While{system running}
    \State Bubble forms with stochastic pressure $P_{\text{air}}(t)+\delta P(t)$
    \State Bubble detaches when $F_b > F_s$
    \State Bubble swarm evolves via Navier–Stokes flow
    \State Optical field $\Psi(\mathbf{r},t)$ evolves through chaotic refraction
    \State  Sensor measures integrated intensity $I(t)$
    \State Convert to 16-bit reading $X_t$
    \State Extract physical bits $B(t)$ from $X_t$
\EndWhile
\end{algorithmic}
\end{algorithm}

\begin{figure}[H]
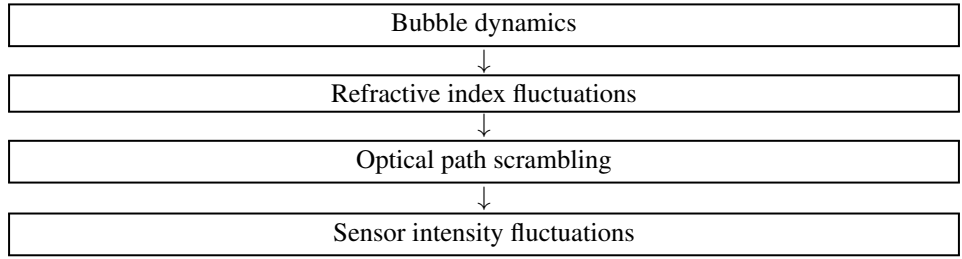

\centering
\vspace{6pt}

\fbox{\parbox{0.7\linewidth}{\centering Bubble dynamics}} \\
$\downarrow$ \\
\fbox{\parbox{0.7\linewidth}{\centering Refractive index fluctuations}} \\
$\downarrow$ \\
\fbox{\parbox{0.7\linewidth}{\centering Optical path scrambling}} \\
$\downarrow$ \\
\fbox{\parbox{0.7\linewidth}{\centering Sensor intensity fluctuations}}\\

\caption{Entropy propagation pipeline in the RLITBW system. Chaotic bubble dynamics induce refractive index fluctuations, which scramble optical paths and result in measurable sensor intensity variations used for random number generation.}
\label{fig:entropy_flow}
\end{figure}

\textbf{Causality of Entropy Generation :}
For clarity, the entropy generation mechanism in the RLITBW system follows a well-defined causal chain. The stochastic formation, deformation, and ascent of air bubbles in the water column induce rapid, spatially non-uniform fluctuations in the local refractive index of the medium. These refractive index fluctuations dynamically scramble the optical propagation paths of the incident light, producing time-varying intensity patterns after refraction through the turbulent fluid. The resulting optical path scrambling is transduced into measurable electrical signals as intensity fluctuations at the light sensor. While each stage of this process is governed by deterministic physical laws, their nonlinear coupling and extreme sensitivity to initial conditions render the observed output effectively unpredictable and practically non-deterministic over time. As shown in Fig.~\ref{fig:entropy_flow}, the entropy in the RLITBW system propagates from fluid dynamics to optical and electronic domains.

\FloatBarrier
\section*{Experimental Setup and Measurement Protocol}
\label{sec:experimental_setup}

This section describes the experimental apparatus, the data-acquisition procedure,
the digital encoding used for storage and transport, and the post-acquisition
processing pipeline used to quantify and distill entropy. The description is
intended to be fully reproducible: all parameters that affect randomness and
statistical evaluation are stated explicitly.

\subsection*{Apparatus and Optical Geometry}

The physical apparatus consists of four functional subsystems: (1) a fluid
container (transparent vessel) in which air bubbles are generated; (2) an
actuated air supply that creates a continuous stream of bubbles through a
controlled orifice; (3) a steady illumination source positioned beneath the
vessel to generate near-parallel light through the water column; and (4) a
single-element light-intensity sensor placed on the opposite side to measure
the time-varying integrated intensity of the transmitted light (BH1750 sensor) \cite{bh1750}. A Raspberry Pi 5 (4GB RAM variant) served as the peripheral device to interface with and control the sensor \cite{raspberrypi}.

The vessel geometry and relative placement of illumination and detector are
selected so that rising bubbles intercept and refract the light path before the
sensor. A relatively dark room was used and it was kept in mind that too much bright light from everywhere will cause the system to produce different bits because the sensor has a particular range of intensity detection. Modifying the source intensity, source-to-vessel distance, or the sensor aperture changes the dynamic range and therefore the effective entropy per sample; all such parameters were tuned empirically to ensure measurable
temporal fluctuation in the recorded intensity.

\begin{figure}[ht]
    \centering
    \includegraphics[width=1\linewidth]{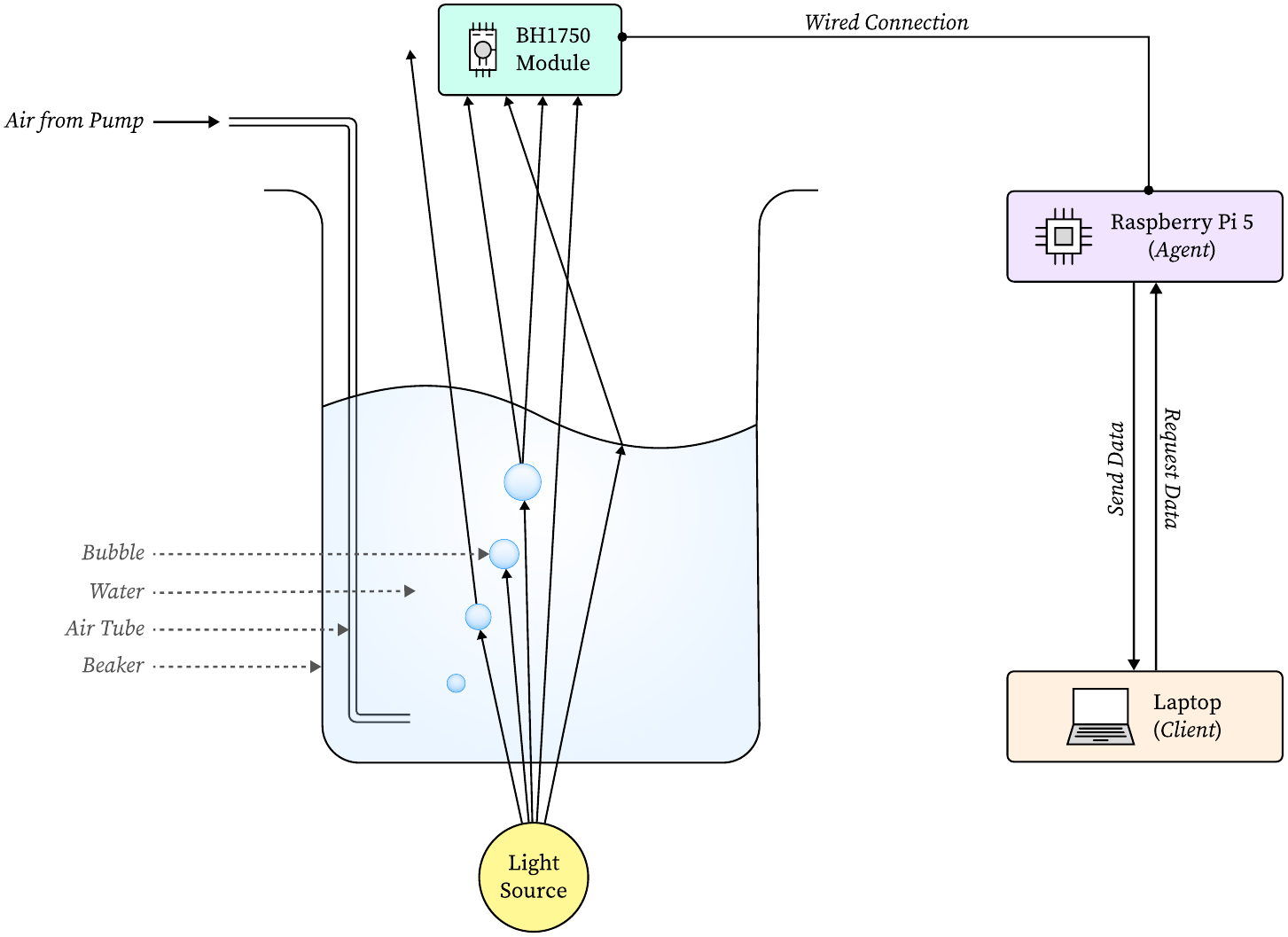}
    \caption{Schematic representation of the experimental setup (illumination,
    bubbling column, and single-element light sensor).}
    \label{fig:setup}
\end{figure}
\begin{figure}[ht]
    \centering
    \includegraphics[width=0.5\linewidth]{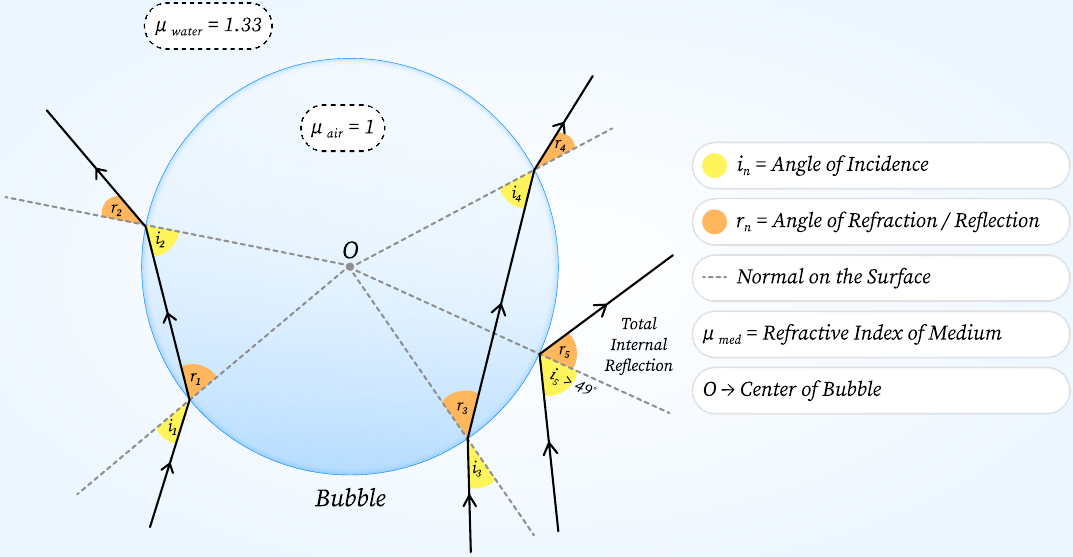}
    \caption{Schematic representation of how a bubble (at an instantaneous time) can alter the directions of incoming light rays from the bottom of the apparatus. Here the bubble is shown as a sphere but in real its surface always moves and thus do not keep a fixed geometry.}
    \label{fig:bubble}
\end{figure}

\subsection*{Fluid and Bubble Parameters}

Key physical parameters used during experiments are:

\begin{itemize}
    \item Orifice diameter: defines detachment threshold and bubble size distribution.
    \item Bulk liquid depth and vessel cross-section: affect bubble rise trajectories.
    \item Air flow (mean pressure and mechanical stability): sets mean bubble creation rate.
    \item Illumination intensity and collimation: determine sensor dynamic range.
\end{itemize}

During data collection the air supply was operated at a fixed nominal setting;
short-term mechanical fluctuations of the pump provide the microscopic pressure
variations that trigger stochastic differences in bubble detachment timing and
initial volumes.

\subsection*{Sensing and Electronic Configuration}

The detector is a single photometric sensor that outputs a quantized intensity
reading at each sampling instant. The acquisition chain includes:

\begin{enumerate}
    \item A synchronous read operation of the sensor yielding an unsigned
    integer sample with fixed bit-depth (16-bit nominal resolution).
    \item A fixed inter-sample interval (empirically chosen in the tens to
    hundreds of milliseconds range) that trades temporal resolution for
    integration-induced mixing of optical speckle patterns.
    \item A lossless integer-to-binary mapping: each quantized sensor value is
    represented as a fixed-length binary word. The ordered concatenation of
    these words constitutes the raw digital bitstream used in subsequent
    processing. ADC nonlinear effects and sampling imperfections can themselves contribute measurable entropy and have been used as practical TRNG sources \cite{adc_nonlinear2021, adc_entropy2020}.
\end{enumerate}

\subsection*{Data Transport and Recording Protocol}

Recorded sensor samples are transferred from the acquisition module to a host
machine using a standard reliable byte-stream protocol. Each sensor sample is
serialized as a two-byte unsigned integer in network byte order and appended to
the host-side buffer. The host decodes each received two-byte word to obtain
the original 16-bit reading and immediately converts it to a fixed-width
binary string (16 characters `0/1'), which is appended to a persistent
log that represents the continuous raw bitstream.

All timing, buffer handling, and framing are implemented so that boundary
misalignment is avoided: bytes are consumed in two-byte chunks and the host
maintains a residual byte buffer to handle partial reads. The acquisition
interval used in experiments was chosen to be larger than the sensor's minimum
integration time to avoid saturating the detector while still capturing
temporal variations induced by bubble motion.

\subsection*{Raw Data Format}

The raw digital sequence is a concatenation of 16-bit binary words:
\[
\underbrace{b_{0}^{(0)} b_{1}^{(0)} \dots b_{15}^{(0)}}_{\text{sample } 0}
\;
\underbrace{b_{0}^{(1)} b_{1}^{(1)} \dots b_{15}^{(1)}}_{\text{sample } 1}
\;
\cdots
\]
where $b_j^{(i)}\in\{0,1\}$ denotes the $j$-th bit (0 = most-significant) of
sample $i$. The representation preserves bit order; this convention is used
consistently during entropy analysis and extraction.

\subsection*{Preprocessing and Statistical Measurement}

Before any extraction or cryptographic expansion, the following preprocessing
and measurement steps are performed:

\begin{enumerate}
    \item \textbf{Truncation to contiguous blocks:} The continuous bitstream is
    partitioned into non-overlapping blocks of fixed sample-count \(S\). In
    reported experiments \(S = 256\) (hence each block contains \(S\cdot 16\)
    raw bits).
    \item \textbf{Basic statistics:} For each block we compute bit-frequency,
    average Hamming weight per sample, and transition count. These empirical
    metrics are used to identify gross bias and temporal correlation.
    \item \textbf{Min-entropy estimation:} A conservative min-entropy per
    sample estimate is computed using symbol-frequency analysis on the integer
    samples. The estimated min-entropy (in bits per 16-bit sample) guides the
    choice of extraction parameters.
\end{enumerate}

\subsection*{Entropy Distillation (Universal Hashing)}

To convert the biased raw samples into a short, high-quality uniform seed we
apply a universal hashing-based extractor with the following characteristics:

\begin{itemize}
    \item \textbf{Input block size:} each extractor invocation consumes a block
    of \(n = S\) samples (here \(n = 256\), i.e., \(n\cdot 16 = 4096\) bits).
    \item \textbf{Security parameter:} the statistical error is bounded by a
    small parameter \(\varepsilon\) (we select \(\varepsilon \le 2^{-40}\) for
    conservatism).
    \item \textbf{Output length:} from each input block a fixed-length uniform
    string of \(k\) bits is produced; in reported experiments \(k = 256\).
\end{itemize}

Mathematically, the extractor implements a linear universal hashing operation
(modulo two) that maps the input vector \(x\in\{0,1\}^{N}\) (with
\(N = 4096\)) to an output \(y\in\{0,1\}^{k}\). The choice of \(k\) respects
the theoretical bound \(k \lesssim H_{\min}(x) - 2\log_2(1/\varepsilon)\),
where \(H_{\min}(x)\) is the estimated min-entropy of the input block.

\subsection*{Cryptographic Expansion (Seeded Stream Generation)}

Each distilled \(k\)-bit output serves as an initial seed for a cryptographic
stream generation primitive. The primitive:

\begin{itemize}
    \item accepts the extracted seed as a secret initialization value,
    \item deterministically produces an arbitrarily long pseudo-random byte
    stream that is computationally indistinguishable from uniform under the
    standard cryptographic assumptions,
    \item is used solely as an \emph{expansion} mechanism – it does not
    increase true entropy beyond what is present in the seed.
\end{itemize}

For evaluation and NIST-style testing we expand some seeds to produce 10 sets of
one million (1\,000\,000) unbiased bits. The expansion output is converted to
ASCII bit representation for compatibility with standard statistical test
batteries (more on it in Section Statistical Evaluation Results). Multiple seeds were used with pseudo-randomly determined expansion limits. Seeds were discarded after full expansion and replaced sequentially. The later process was used in the optimization algorithms (in Section Some Application to Optimization Algorithms) so as to produce a large amount of random numbers faster.

\subsection*{Reproducibility, Deterministic Seeding, and Publication}

The universal hashing procedure uses a public deterministic matrix generation
method (derived from a cryptographic hash-chain) so that experimenters can
reconstruct the exact extractor used from the published description. The
stream-generation primitive is also instantiated in a deterministic,
documented manner (fixed initialization vector) to enable full reproducibility.
Publication of the raw dataset, the extractor matrix specification, and the
seed-to-stream mapping is recommended to allow independent verification.

\subsection*{Practical Parameter Choices and Rationale}

For transparency, the principal parameter choices used in reported experiments
were:

\begin{itemize}
    \item Sampling interval: selected to allow sensor integration while
    preserving bubble-induced variation (order of 0.1–0.2 seconds).
    \item Block size for extraction: \(S = 256\) samples (4096 input bits).
    \item Extracted seed length: \(k = 256\) bits per block.
    \item Expansion target: 1\,000\,000 output bits per seed.
    \item Security margin: choose \(\varepsilon \le 2^{-40}\) when calculating
    maximum safe extraction length per block.
\end{itemize}

These values were chosen to balance statistical safety, runtime convenience,
and the limited amount of physically-obtained raw data available in one
experimental run.

\subsection*{Operational Notes}

\begin{itemize}
    \item The expansion stage is cryptographically secure only if the seed
    contains sufficient true entropy. The extraction stage must therefore be
    described in the methods for any security claim.
    \item Environmental changes (illumination drift, pump speed drift,
    temperature) can alter per-sample entropy; therefore each experimental run
    should report contemporaneous environmental metadata. But reproducibility is not guaranteed.
    \item For strict randomness certification it is preferable to publish both
    (a) raw acquisition logs and (b) the post-extraction and expanded bit-sets
    used in statistical tests so readers can reproduce the analysis.
\end{itemize}

\newpage
\subsection*{Algorithmic Summary}

\begin{algorithm}[htbp]
\caption{Measurement and Processing Pipeline (conceptual)}
\begin{algorithmic}[1]
\State Initialize apparatus and stabilize illumination and pump
\While{acquisition desired}
    \State Acquire sample $X_t$ at inter-sample interval $\Delta$
    \State Serialize $X_t$ as fixed-width binary
    \State Append to raw stream
    \If{raw stream length $\ge$ block size}
        \State Partition block into vector $x\in\{0,1\}^N$
        \State Estimate $H_{\min}(x)$, verify extraction preconditions
        \State Compute $y = \mathrm{Extractor}(x)$, $|y|=k$
        \State Produce expanded stream $\mathrm{Expand}(y)$ as needed
        \State Log raw block, $y$, and expansion metadata
    \EndIf
\EndWhile
\end{algorithmic}
\end{algorithm}


\section*{Entropy Extraction, Conditioning, and Bitstream Expansion}
The physical system described in the previous section produces a discrete-time
sequence of 16-bit observations $\{X_t\}$ obtained from the photometric
integration of the dynamic refractive field above the bubbling water column.
Although the underlying dynamics exhibit strong chaotic behavior, the resulting
raw digital samples generally contain bias, correlations, and
non-uniform symbol probabilities due to sensor characteristics, finite optics,
and turbulent structures with memory. Consequently, a post-processing pipeline
is required to (i) isolate the physically generated entropy, (ii) compress it
into a shorter but higher-quality seed, and (iii) expand the seed into a
longer bitstream suitable for statistical evaluation and reproducible
distribution.

In this work, the post-processing architecture follows the common structure
prescribed by cryptographically secure TRNG conditioning standards such as
NIST SP~800-90B~\cite{nist80090b} and SP~800-22~\cite{nist80022}. The approach
consists of three stages: (1) raw-bit acquisition, (2) entropy
extraction using a linear universal hashing operator, and (3) deterministic
expansion using a well-studied stream-generation primitive.
Most important codes and data found in the study is made public in github \cite{github_repo}.

\subsection*{Acquisition of Raw Physical Bits}

Each photometric sample $X_t$ is a 16-bit integer corresponding to the optical
intensity integrated over the sensor area and exposure window. To obtain a
binary sequence suitable for subsequent conditioning, every sample is
represented as a fixed-length binary word:

\begin{equation}
    X_t = \sum_{k=0}^{15} x_{t,k} 2^k, \qquad x_{t,k} \in \{0,1\},
\end{equation}

and the raw bitstream is defined as

\begin{equation}
    R = (x_{0,15}, x_{0,14}, \ldots, x_{0,0},\;
         x_{1,15}, \ldots ).
\end{equation}

Since lower-order bits typically carry greater physical entropy in analog–digital
measurements, the extracted sequence contains the contributions of both
high-entropy components and low-entropy components, necessitating statistical
compression.

Given $N$ collected samples, the initial dataset contains $16N$ raw bits.
For the entropy extractor used in the NIST testing, the first $4096$ bits (corresponding
to $256$ consecutive sensor readings) are taken as the input block
\[
    R_{\text{block}} \in \{0,1\}^{4096}.
\]
Later, we set the expansion limit pseudo-randomly for including more entropy collected from the setup.

\subsection*{Entropy Source Characterization}
Before post-processing, the raw 16-bit integer data stream from the light intensity sensor was analyzed to validate the chaotic nature of the physical entropy source. The turbulent flow of air bubbles through water creates a non-periodic and unpredictable fluctuation in the refracted light intensity, which is captured by the sensor.

\begin{figure*}[ht]
    \centering
    \includegraphics[width=\linewidth]{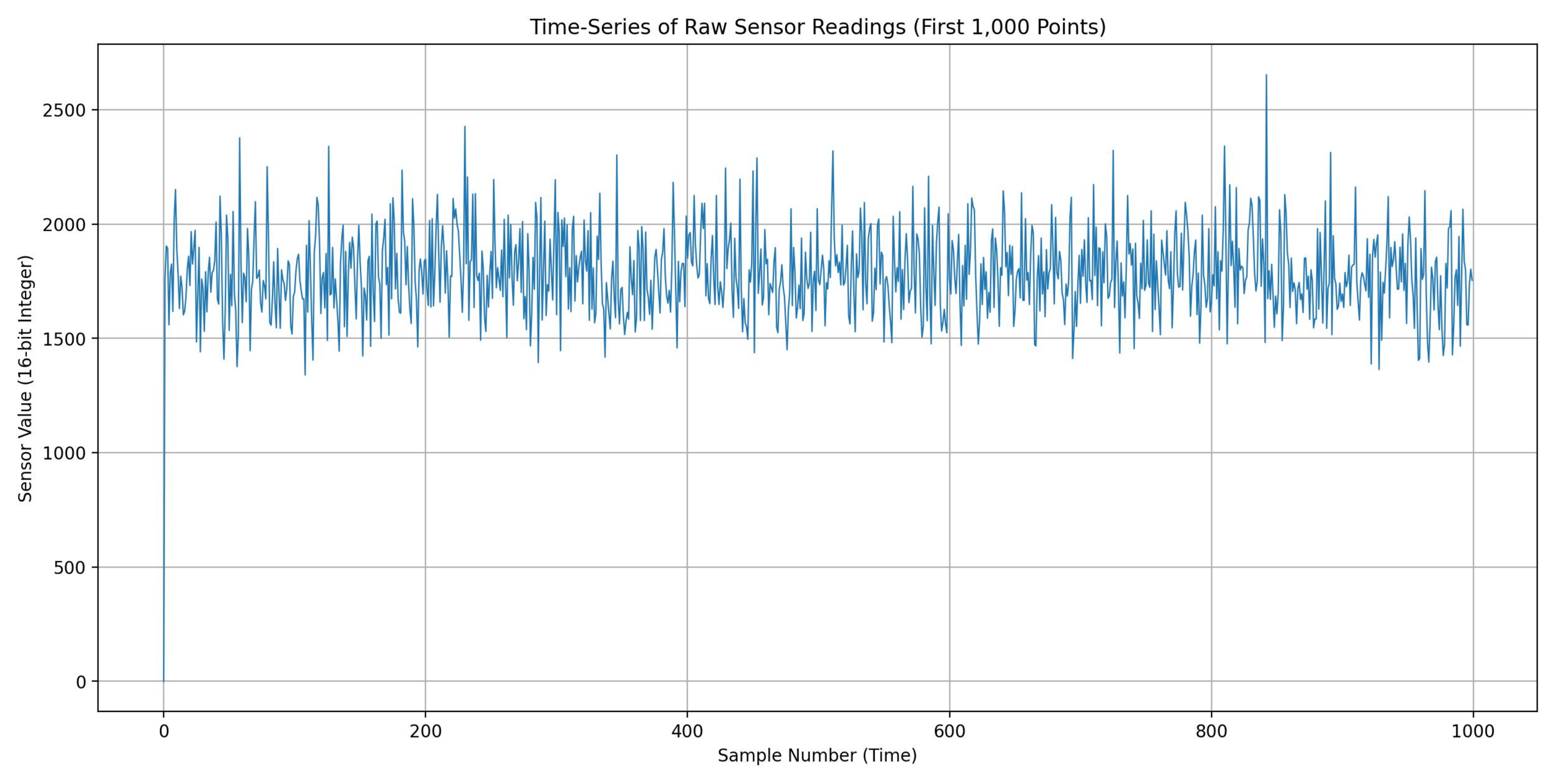} 
    \caption{Time-series of raw light intensity measurements captured from the RLITBW system. The signal exhibits irregular, aperiodic fluctuations arising from the combined effects of turbulent bubble dynamics and optical path scrambling, forming the basis for subsequent entropy extraction and statistical validation.}

    \label{fig:time_series}
\end{figure*}

The time-series plot in Fig. \ref{fig:time_series} shows erratic, non-repeating fluctuations around a stable mean, which is the expected visual signature of a chaotic process.

\begin{figure}[ht]
    \centering
    \includegraphics[width=0.5\linewidth]{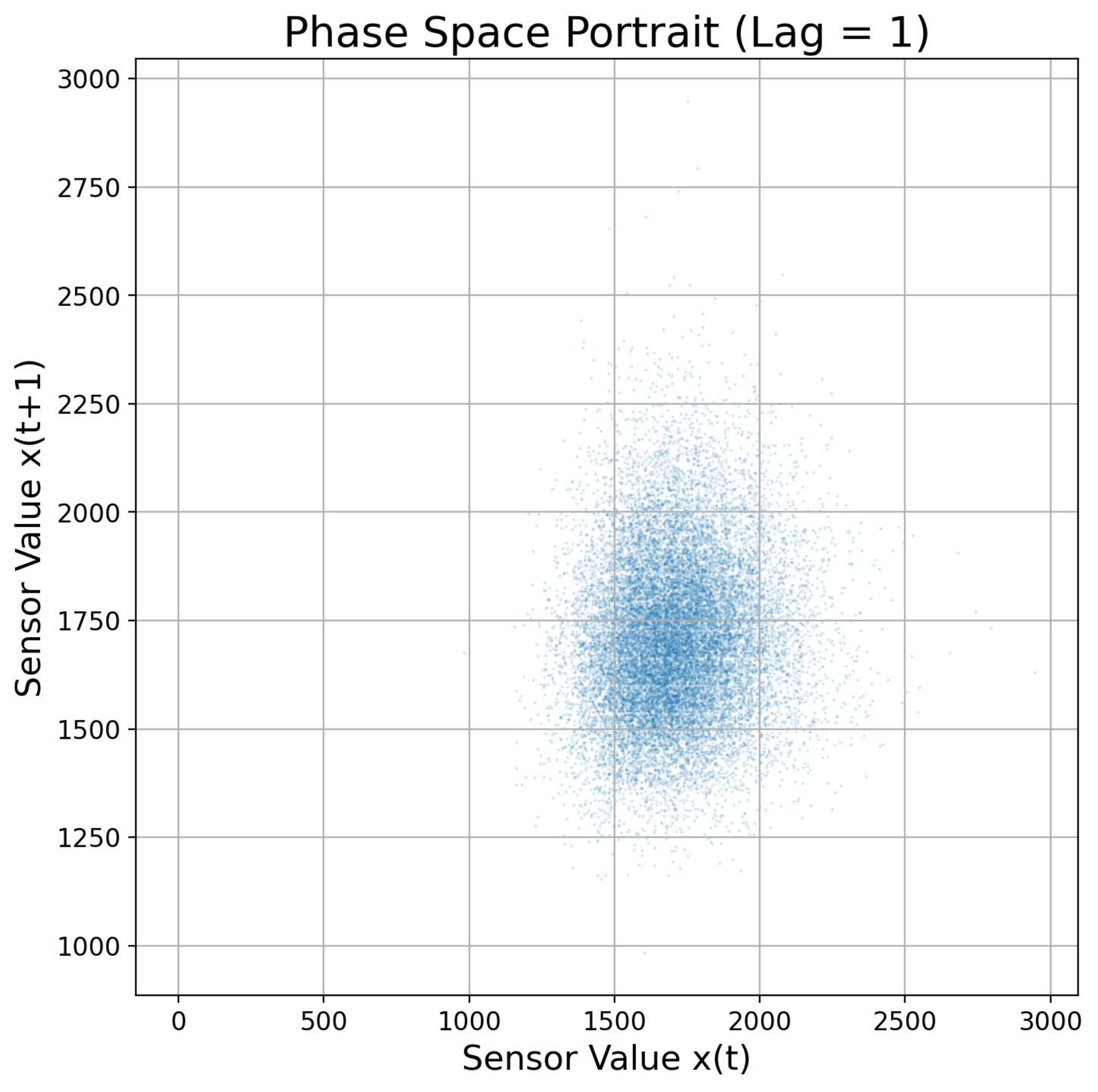} 
    \caption{Phase-space reconstruction of the RLITBW sensor signal using time-delay embedding. The plot shows the sensor intensity $x(t)$ versus its delayed version $x(t+\tau)$, where $\tau$ is the embedding delay. The resulting non-periodic, structured geometry indicates deterministic chaotic dynamics with sensitive dependence on initial conditions, supporting the suitability of the underlying physical process as an entropy source.}

    \label{fig:phase_space}
\end{figure}

To further investigate the underlying dynamics, a phase space portrait \cite{takens1981} was constructed (Fig. \ref{fig:phase_space}). The emergence of a cloud of dots provides strong evidence of chaos because a non-uniform Independent and Identically Distributed sequence produces such kind of plots at delay=1 (or Lag=1).
We also checked different Lags={1, 2, 3, 5, 10} but all of them resulted in similar structures.

\begin{figure}[ht]
    \centering
    \includegraphics[width=0.5\linewidth]{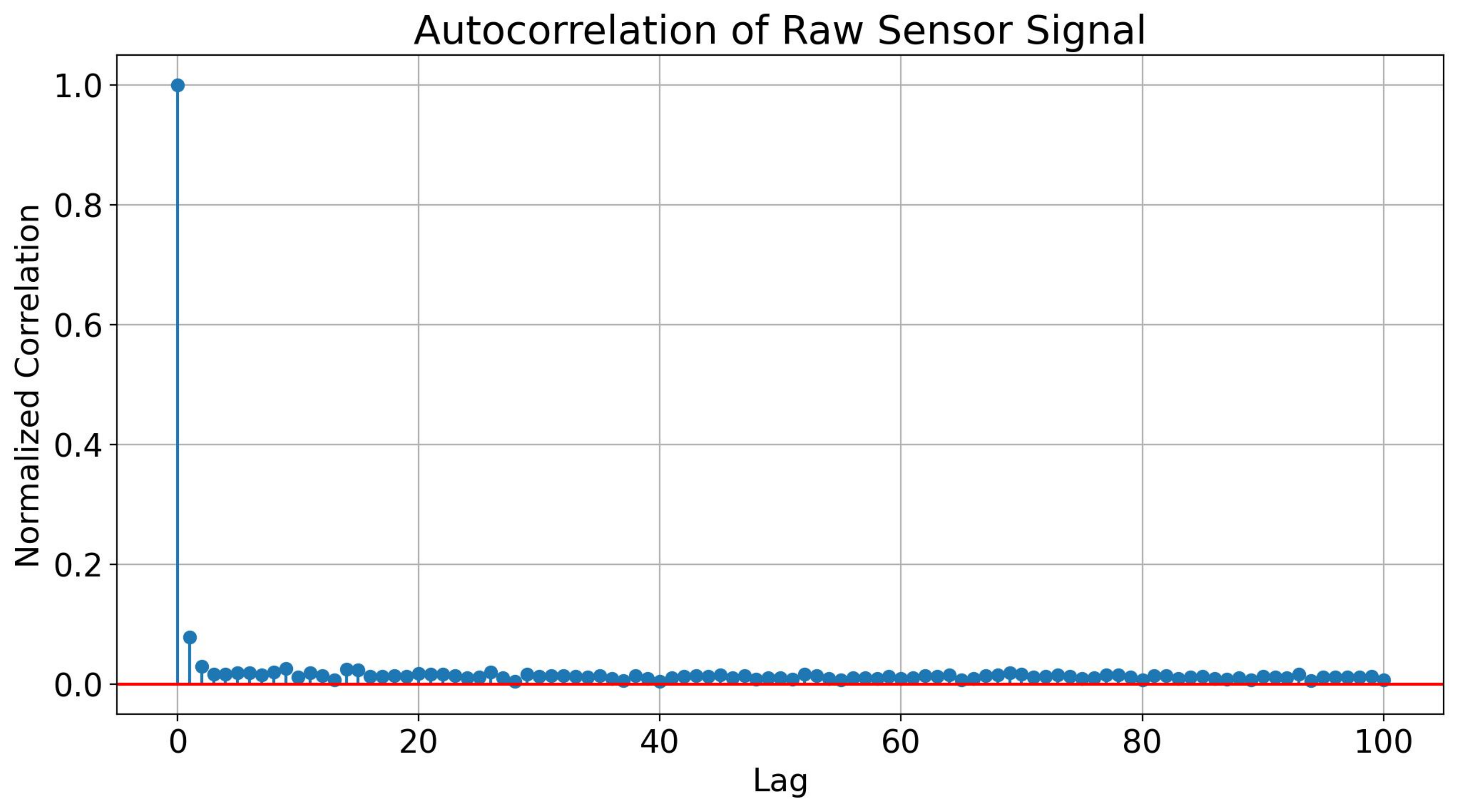} 
    \caption{Autocorrelation of the raw sensor signal, showing a rapid drop to zero, indicating high unpredictability.}
    \label{fig:autocorrelation}
\end{figure}

The unpredictability of the signal was quantified using autocorrelation analysis (Fig. \ref{fig:autocorrelation}). The correlation function drops to a statistically insignificant value immediately at lag=1, confirming that the system has a very short memory and successive readings are highly independent.

Finally, the dynamical behavior of the RLITBW time series was examined using the largest Lyapunov exponent estimated from the reconstructed phase space following the method of Wolf et al. \cite{wolf1985}. A small but positive value \textbf{\(\lambda = \textbf{0.0036}\)} was obtained, indicating divergence of nearby trajectories under identical reconstruction assumptions and reflecting the highly irregular nature of the signal. Due to the presence of inherent physical randomness, this Lyapunov exponent is interpreted as a dynamical indicator rather than a strict measure or proof of deterministic chaos.

In addition, the complexity of the signal was quantified using Sample Entropy \cite{richman2000}, which yielded a high value of 2.1333, signifying strong unpredictability and low temporal regularity. Taken together, these measures support the characterization of the RLITBW output as a physically generated entropy source \cite{minentropy_ssl2022}.


\subsection*{Linear Hash-Based Entropy Extraction}

To distill the inherent physical randomness contained in $R_{\text{block}}$,
we apply a linear, universal hash family in the form of a Toeplitz matrix
compression operator. Methods for entropy evaluation and Toeplitz/universal hashing postprocessing are well established for optical RNGs \cite{ma2013toeplitz}. Toeplitz hashing is widely used in information-theoretic
entropy extractors~\cite{dodismiller}, \cite{ma2013toeplitz} because it provides:

\begin{itemize}
    \item provable universality,
    \item strong resilience to bias and linear correlations,
    \item efficient hardware and software implementation,
    \item suitability for min-entropy–limited sources.
\end{itemize}

\subsubsection*{Toeplitz Matrix Definition}

A Toeplitz matrix $T \in \{0,1\}^{m \times n}$ is defined by a sequence of
$m+n-1$ bits:

\begin{equation}
    T =
    \begin{bmatrix}
    t_0     & t_1     & t_2     & \cdots & t_{n-1} \\
    t_{-1}  & t_0     & t_1     & \cdots & t_{n-2} \\
    t_{-2}  & t_{-1}  & t_0     & \cdots & t_{n-3} \\
    \vdots  & \vdots  & \vdots  & \ddots & \vdots \\
    t_{-(m-1)} & t_{-(m-2)} & \cdots & \cdots & t_0
    \end{bmatrix},
\end{equation}

where each descending diagonal is constant. This structure ensures that the
matrix can be fully parameterized by $m+n-1$ seed bits.

\subsubsection*{Entropy Compression Step}

The extractor maps $R_{\text{block}}$ to a shorter vector

\begin{equation}
    S = T \cdot R_{\text{block}} \pmod{2},
\end{equation}

where $S \in \{0,1\}^{256}$ is a 256-bit seed.

The compression (4096 raw bits $\rightarrow$ 256 conditioned bits)
concentrates the physical entropy under the assumption that the input block
contains at least 256 bits of min-entropy. Such a compression is consistent with
the Leftover Hash Lemma~\cite{hastad1999}, which guaranties statistical
indistinguishability from uniform when the source has sufficient min-entropy.

The Toeplitz parameters used here are deterministically generated from a
public initialization value to ensure reproducibility for independent
evaluation, yet they remain independent of the TRNG output, satisfying the
requirements for a strong extractor seed.

\subsection*{Deterministic Expansion via Stream Generation}

After the 256-bit seed $S$ is obtained, it is treated as the state of a
deterministic pseudorandom expansion function. This step does not increase
entropy but rather spreads the extracted entropy uniformly across a longer
output sequence, enabling convenient testing using large-scale randomness
suites.

Let the expansion function be denoted by

\begin{equation}
    G : \{0,1\}^{256} \rightarrow \{0,1\}^M,
\end{equation}

where $M = 10^6$ bits for the evaluations in this work. The generator produces
a keystream

\[
    Y = G(S)
\]

which is interpreted as an unbiased surrogate for the measured entropy rate.
This methodology follows standard practice in conditioning-compliant TRNG
design, where a small, high-quality physical seed is expanded to a longer
sequence for statistical validation. Because the generator is deterministic,
all unpredictability originates solely from the physical system.

\subsection*{Minimum Entropy Analysis}
Min-entropy analysis was performed to quantify the worst-case unpredictability of the proposed RLITBW-based entropy source. Each sensor measurement was digitized as a 16-bit sample. For the raw sensor output, analysis of 1,000,128 bits (62,508 samples) yielded a min-entropy of 8.37 bits per sample, corresponding to 0.52 bits per bit, indicating the presence of significant physical entropy along with measurable bias and temporal correlation. Block-level analysis over four consecutive samples further reduced the estimated min-entropy to 3.48 bits per sample (0.22 bits per bit), motivating the need for entropy conditioning. Subsequently, Toeplitz hashing followed by ChaCha20 was applied as a deterministic randomness extractor to redistribute the available entropy without introducing additional randomness. Post-processed output, evaluated over 10,000,000 bits (625,000 samples), exhibited an improved min-entropy of 14.61 bits per 16-bit sample, corresponding to 0.91 bits per bit, with a block min-entropy of 4.31 bits per sample (0.27 bits per bit), demonstrating effective reduction of bias and correlation and suitability for cryptographic use. Sensor noise under constant illumination is minimal (±2 lux). However, light intensity variations within the sensor's dynamic range contribute measurably to entropy.

\subsection*{Reasons For Toeplitz Hashing}

The combined Toeplitz-extraction and deterministic expansion pipeline offers a
transparent, mathematically analyzable approach to separating physical entropy
from deterministic processing. The Toeplitz stage isolates the genuine
randomness inherent in the optical-fluidic system, while the expansion stage
allows generation of arbitrarily long sequences for standard randomness tests
without requiring prohibitively long experimental acquisition periods. This
structure is commonly adopted in high-assurance TRNG designs where physical
sampling may be slow, costly, or fragile.


\section*{Statistical Evaluation Results}
\label{sec:section_vi}
After verifying the chaotic nature of the raw physical source, the final 1-megabit output stream (post-Toeplitz extraction and ChaCha20 \cite{bernstein2008chacha} deterministic expansion) was subjected to a comprehensive set of statistical evaluations to confirm its quality and suitability for cryptographic applications.

\subsection*{Fundamental Randomness and Entropy Checks}
As a preliminary step, two fundamental tests were performed to verify the basic properties of randomness.

First, a frequency test (monobit) was conducted to measure statistical bias. Out of 1,000,000 bits, the distribution was exceptionally well-balanced:
\begin{itemize}
    \item \textbf{Count of '0's:} 499,858 (49.99\%)
    \item \textbf{Count of '1's:} 500,142 (50.01\%)
\end{itemize}
This near-perfect 50/50 distribution confirms that the output stream is unbiased.

Second, to assess the entropy of the bitstream, an optional lossless compression (zlib) was applied. The data was found to be incompressible, producing a compression ratio of 1.25, where the compressed file was 25\% larger than the original due to overhead. This result strongly indicates high entropy and a lack of discernible patterns, which is the expected behavior of a secure random sequence.

\subsection*{NIST SP 800-22 Test Suite Evaluation}
Following the fundamental checks, the generated bitstream was subjected to the full NIST SP 800-22 Rev 1a test suite. The suite consists of 15 distinct statistical tests designed to detect non-randomness in binary sequences. For this evaluation, a sequence length of $10^6$ bits was used. The results, summarized in Table \ref{tab:nist_results}, demonstrate that the generator passed all applicable tests, providing strong evidence of its randomness quality. In addition to NIST SP 800-22, empirical RNG testing frameworks such as TestU01 provide extensive test batteries and were considered (although no results are shown in this study) for supplementary checks \cite{testu012007,testu01_guide2007}.

\subsection*{Tabulated NIST Test Results}
Table \ref{tab:nist_results} presents the breakdown of the test outcomes. The "Proportion" column indicates the pass rate for the sub-tests within each category. For the \textit{Non-Overlapping Template} test, which consists of 148 individual sub-tests, and for the \textit{Cumulative Sums} and \textit{Serial} tests, the results have been aggregated to provide a clear summary.

All tests passed successfully, meeting the criteria for randomness. For the \textit{Random Excursions} and \textit{Random Excursions Variant} tests, a final P-value is not generated. Instead, the assessment is based on the proportion of subsequences passing, which was sufficient in this case. The occasional failure of individual sequences to generate enough cycles for these specific sub-tests is a known property of the suite and does not indicate a failure of the generator itself.

\begin{table*}[ht]
	\centering
	\caption{NIST SP 800-22 Test Results for 10x 1M Bit Sequences}
	\label{tab:nist_results}
	\begin{tabular}{@{}lccc@{}}
		\toprule
		\textbf{Statistical Test} & \textbf{P-VALUE} & \textbf{PROPORTION} & \textbf{Result} \\
		\midrule
		Frequency                    & 0.739918 & 10/10 & \textbf{PASS} \\
		BlockFrequency               & 0.350485 & 10/10 & \textbf{PASS} \\
		CumulativeSums (Forward)     & 0.534146 & 10/10 & \textbf{PASS} \\
		CumulativeSums (Backward)    & 0.350485 & 10/10 & \textbf{PASS} \\
		Runs                         & 0.122325 & 10/10 & \textbf{PASS} \\
		LongestRun                   & 0.739918 & 10/10 & \textbf{PASS} \\
		Rank                         & 0.911413 & 10/10 & \textbf{PASS} \\
		FFT                          & 0.350485 & 10/10 & \textbf{PASS} \\
		NonOverlappingTemplate (148 tests) & 0.000199--0.911413 & 9/10--10/10 & \textbf{PASS} \\
		OverlappingTemplate          & 0.122325 & 10/10 & \textbf{PASS} \\
		Universal                    & 0.122325 & 10/10 & \textbf{PASS} \\
		ApproximateEntropy           & 0.122325 & 10/10 & \textbf{PASS} \\
		RandomExcursions (8 tests)   & ---      & 5/5   & \textbf{PASS} \\
		RandomExcursionsVariant (18 tests) & ---  & 5/5   & \textbf{PASS} \\
		Serial (2 tests)             & 0.035174--0.534146 & 10/10 & \textbf{PASS} \\
		LinearComplexity             & 0.739918 & 10/10 & \textbf{PASS} \\
		\bottomrule
	\end{tabular}
	\begin{minipage}{\textwidth}
		\vspace{0.2cm}
		\small
		\textit{Note:} The minimum pass rate for each statistical test (except random excursion variants) is approximately 8 for a sample size of 10 binary sequences. For random excursion (variant) tests, the minimum pass rate is approximately 4 for a sample size of 5 binary sequences.
	\end{minipage}
\end{table*}

\subsection*{Graphical Analysis of NIST P-Values}
To further validate the uniformity and quality of the generator, we analyze the distribution of P-values produced by the NIST tests. For a truly random source, the P-values from these statistical tests should themselves be uniformly distributed between 0 and 1.

\begin{figure}[htbp]
	\centering
	\includegraphics[width=0.5\linewidth]{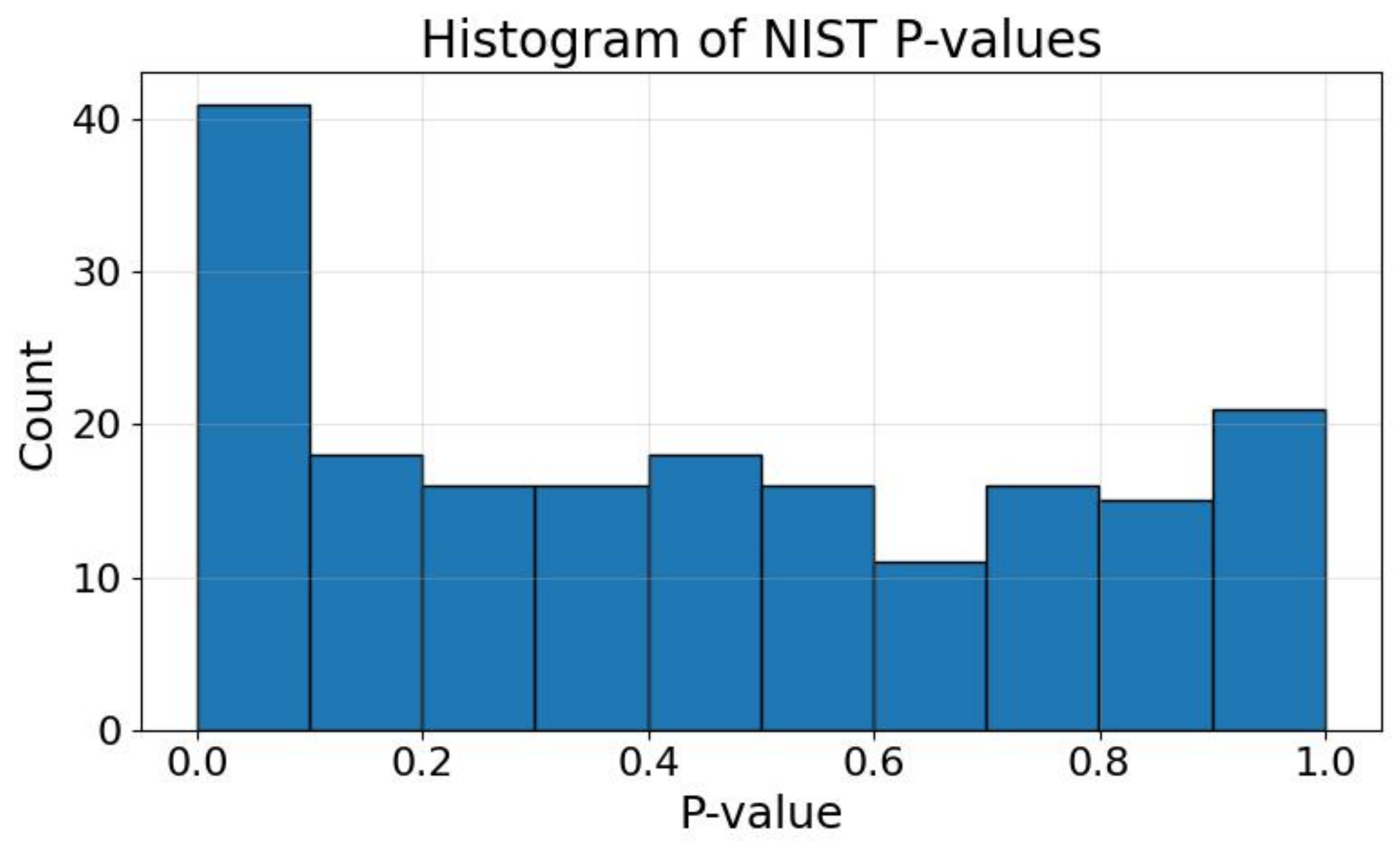} 
	\caption{Histogram showing the frequency distribution of P-values from the NIST tests. The relatively flat and uniform spread is consistent with the behavior of a true random source.}
	\label{fig:p_value_histogram}
\end{figure}

\begin{figure}[ht]
\centering

\begin{subfigure}{0.48\linewidth}
    \centering
    \includegraphics[width=\linewidth]{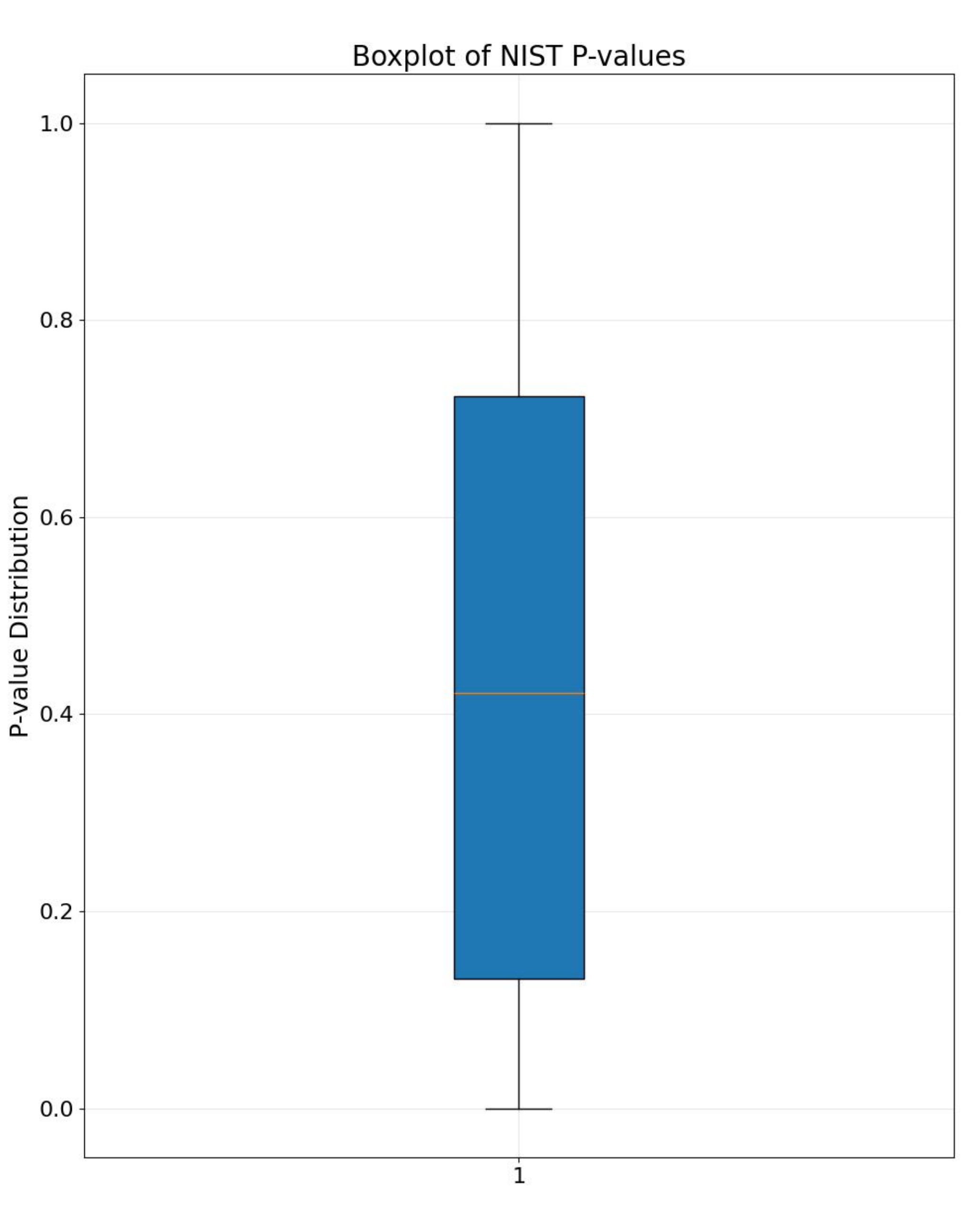}
    \caption{Boxplot analysis of the P-value distribution. The median is close to the expected 0.5, and the interquartile range shows a healthy spread without significant outliers, supporting uniformity.}
    \label{fig:p_value_boxplot}
\end{subfigure}
\hfill
\begin{subfigure}{0.48\linewidth}
    \centering
    \includegraphics[width=\linewidth]{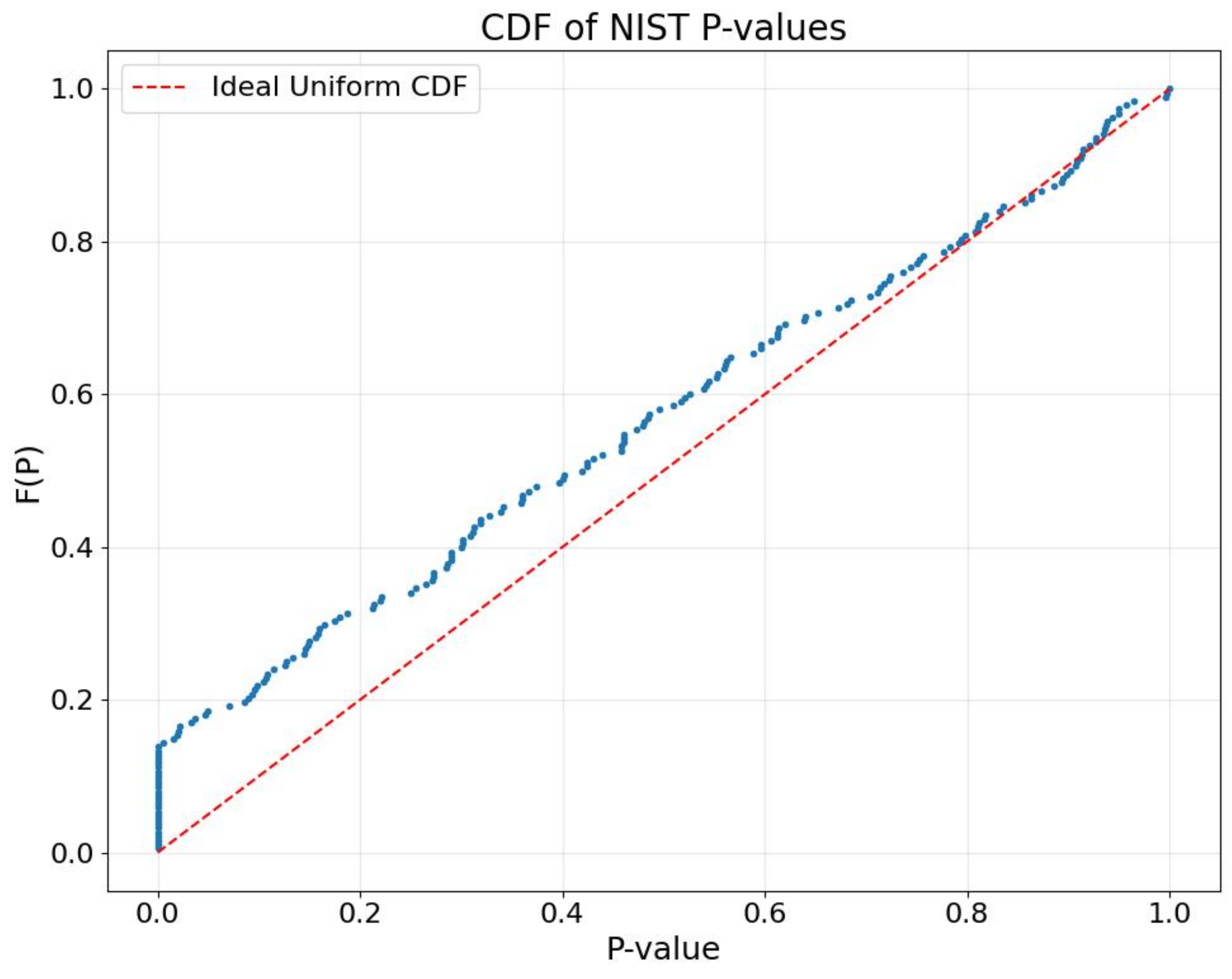}
    \caption{CDF of P-values vs ideal uniform CDF (dashed red line). The close adherence demonstrates agreement with the expected uniform distribution.}
    \label{fig:p_value_cdf}
\end{subfigure}

\caption{Statistical validation of randomness using P-value distribution analysis.}
\label{fig:p_value_analysis}

\end{figure}

Figures \ref{fig:p_value_histogram}, \ref{fig:p_value_cdf} illustrate the histogram and Cumulative Distribution Function (CDF) of the P-values. The histogram displays a spread consistent with uniformity, and the CDF closely tracks the ideal $y=x$ diagonal, indicating that there is no systematic bias in the test scores that would suggest non-randomness.

\begin{figure}[ht]
	\centering
	\includegraphics[width=0.5\linewidth]{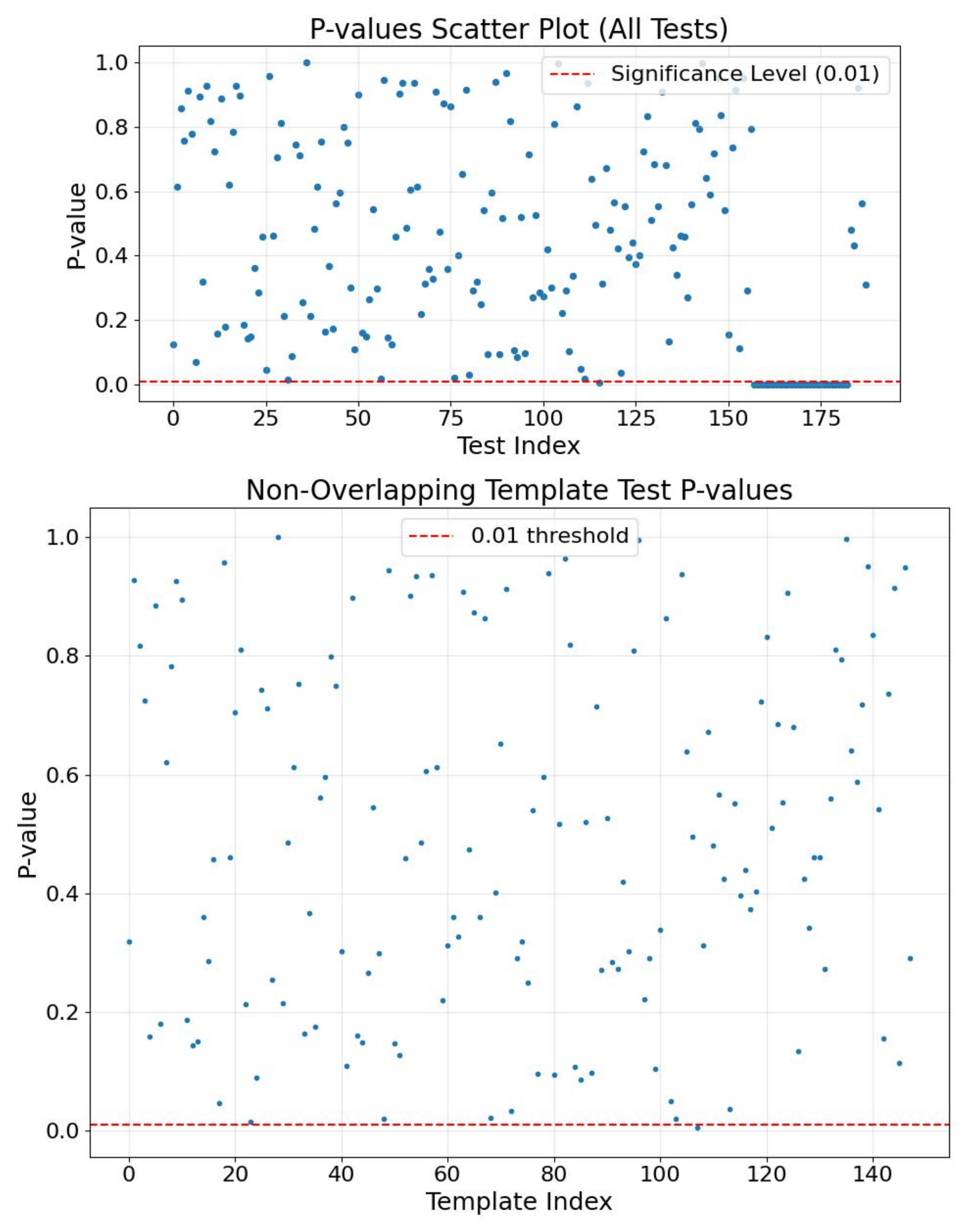}
	\caption{Scatter plot visualization of test results. Top: P-values for all executed tests indexed sequentially; all points lie above the $\alpha=0.01$ failure threshold. Bottom: Detailed scatter plot for the 148 Non-Overlapping Template sub-tests, demonstrating a random spread of P-values.}
	\label{fig:nist_scatter}
\end{figure}

Figure \ref{fig:nist_scatter} provides a scatter plot of the P-values for all executed tests. All recorded values fall well above the significance threshold ($\alpha = 0.01$), represented by the dashed red line. The bottom pane specifically highlights the 148 sub-tests of the Non-Overlapping Template test, showing a healthy variance without clustering near the failure threshold, further confirming the robustness of the output.

\section*{Some Application to Optimization Algorithms}
\label{sec:section_vii}

To show that the proposed RLITBW TRNG can also be used like a software PRNG in a
real optimization task, we ran three well-known continuous optimization
algorithms on simple benchmark functions using two different randomness sources:
(1) the default PRNG of the implementation and (2) random numbers generated
from the RLITBW bitstream. 

\subsection*{Algorithms and Randomness Sources}

We considered three population-based algorithms that are widely used for
continuous black-box optimization. 

\textbf{CMA-ES} (Covariance Matrix Adaptation Evolution Strategy) \cite{hansen2001cmaes} is an
evolutionary algorithm that samples candidate solutions from a multivariate
normal distribution and then adapts the mean and covariance matrix based on the
best samples in each generation. Random numbers are used whenever new samples
are drawn from the search distribution and when parents are selected. 

\textbf{L-SHADE} (Success-History based Adaptive Differential Evolution with
Linear Population Size Reduction) \cite{tanabe2014lshade} is an improved Differential Evolution method
where the mutation and crossover parameters are adapted from the history of
successful moves, while the population size is slowly reduced over time. The
algorithm calls the random number generator when it selects parents and decides
crossover points for each individual. 

\textbf{SPSO} \cite{kennedy1995pso} is a particle swarm optimization variant where a swarm of
particles moves through the search space. Each particle updates its velocity
using a combination of inertia, attraction towards its own best position and
the global best position, all multiplied by random coefficients. This makes the
quality of the random numbers important for avoiding premature convergence. 

For each algorithm we created two versions. In the baseline version, all random
calls used the standard PRNG known as the Mersenne Twister \cite{mersenne}. In the TRNG-based
version, all random numbers were drawn from the RLITBW output stream, mapped to
floating-point values in the interval \([0,1)\) and then used wherever the
algorithm expects uniform random numbers. Also as mentioned before the creation of the true random numbers was done using multiple 256-bits of seed and expanding them pseudo randomly to different expansion limits for better inclusion of the collected entropy. Both versions used the same general settings so that only the randomness source was changed. 

\subsection*{Benchmark Functions and Experimental Setup}

As a first application study we used four standard continuous test functions:
Ackley, Rastrigin, Rosenbrock and Sphere. These are simple but representative
benchmarks that cover multimodal landscapes with many local minima as well as
smooth convex basins. All functions were treated as minimisation problems. 

For every combination of function and algorithm, both the PRNG-based and the
TRNG-based versions were run several times with independent seeds. We recorded
the final objective value reached at the end of each run and then computed the
mean and standard deviation across runs. Lower mean values indicate better
optimization performance, while the standard deviation shows the stability of
the algorithm across different random seeds. 

\subsection*{Results and Discussion}

Table~\ref{tab:opt_results} summarises the results for all algorithms and
functions. For the Ackley and Sphere functions, both the PRNG and TRNG versions
of CMA-ES reach values extremely close to zero, and the differences are far
below the reported standard deviations, which means that replacing the PRNG
with RLITBW noise does not harm performance. 

For L-SHADE and SPSO the behaviour is also very similar between PRNG and TRNG
runs. On some functions (for example, Ackley with L-SHADE) the TRNG version is
slightly better, while on others (such as Rastrigin with L-SHADE) the PRNG
version has a slightly smaller mean error. In all cases the differences are
small compared to the natural run-to-run variation captured by the standard
deviations. 

These preliminary results demonstrate that the RLITBW TRNG can be
used as a drop-in replacement for a software PRNG in typical metaheuristic
optimisation algorithms without degrading optimisation quality. This gives an
initial demonstration that the proposed entropy source is not only
statistically sound but also practically usable in optimisation workloads. 


\begin{table*}[t]
\centering
\caption{Combined optimization results for all benchmark functions and algorithms. Each entry reports the mean and standard deviation of the final objective value over multiple independent runs (lower is better).}
\label{tab:opt_results}

\setlength{\tabcolsep}{4pt} 
\renewcommand{\arraystretch}{1.15}

\small
\begin{tabularx}{\textwidth}{@{}l l X r r r r@{}}
\toprule
\textbf{Function} 
& \textbf{Algorithm} 
& \textbf{Description} 
& \multicolumn{2}{c}{\textbf{PRNG}} 
& \multicolumn{2}{c}{\textbf{TRNG}} \\
\cmidrule(lr){4-5} \cmidrule(lr){6-7}
& & & \textbf{Mean} & \textbf{Std} & \textbf{Mean} & \textbf{Std} \\
\midrule

\multirow{3}{*}{Ackley}
& CMA-ES  
& Covariance Matrix Adaptation Evolution Strategy
& 3.99e--15 & 0.00e+00 & 4.11e--15 & 6.37e--16 \\
& L-SHADE 
& Success-History Adaptive Differential Evolution
& 8.27e--03 & 4.02e--03 & 7.41e--03 & 3.99e--03 \\
& SPSO    
& Standard Particle Swarm Optimization
& 1.398 & 1.033 & 1.570 & 0.936 \\
\midrule

\multirow{3}{*}{Rastrigin}
& CMA-ES  
& Covariance Matrix Adaptation Evolution Strategy
& 6.268 & 4.322 & 6.467 & 4.919 \\
& L-SHADE 
& Success-History Adaptive Differential Evolution
& 3.389 & 1.461 & 7.766 & 2.589 \\
& SPSO    
& Standard Particle Swarm Optimization
& 17.946 & 7.656 & 21.193 & 10.199 \\
\midrule

\multirow{3}{*}{Rosenbrock}
& CMA-ES  
& Covariance Matrix Adaptation Evolution Strategy
& 6.54e--09 & 3.52e--08 & 1.33e--01 & 7.16e--01 \\
& L-SHADE 
& Success-History Adaptive Differential Evolution
& 7.48 & 0.578 & 7.44 & 0.538 \\
& SPSO    
& Standard Particle Swarm Optimization
& 5.725 & 2.487 & 7.396 & 2.797 \\
\midrule

\multirow{3}{*}{Sphere}
& CMA-ES  
& Covariance Matrix Adaptation Evolution Strategy
& 6.83e--66 & 9.86e--66 & 7.57e--66 & 1.67e--65 \\
& L-SHADE 
& Success-History Adaptive Differential Evolution
& 4.02e--05 & 9.68e--05 & 2.59e--05 & 2.20e--05 \\
& SPSO    
& Standard Particle Swarm Optimization
& 1.59e--06 & 7.59e--06 & 6.12e--06 & 3.00e--05 \\
\bottomrule
\end{tabularx}
\end{table*}

\section*{Practical Deployment and Scalability}
\label{sec:deployment}

The RLITBW entropy source can be deployed as a small, standalone hardware module that continuously generates physical randomness and delivers it to software systems through a buffered interface. Figure~\ref{fig:rlitbw-deployment} provides a conceptual view of this pipeline: the RLITBW device produces raw optical--fluidic measurements, a local controller performs extraction, and the resulting high-quality seeds are stored in an entropy buffer that clients can access on demand. This section presents a conceptual deployment plan; the present work validates the entropy source and processing pipeline, while the full appliance/server architecture is left for future engineering. Alternative compact TRNG technologies such as spintronic devices were studied which offered different throughput/energy trade-offs \cite{spintronics_review2021}

In a minimal single-host deployment, the RLITBW module operates as a dedicated peripheral connected to a microcontroller or a single-board computer. This controller samples the sensor, applies the Toeplitz-based extractor described in Section Experimental Setup and Measurement Protocol, and stores the resulting 256-bit seeds inside a circular in-memory buffer (or a small persistent store). When a client requests random bits, the host consumes one or more stored seeds and expands each seed on demand using a deterministic generator such as ChaCha20 (or other approved primitives depending on the application). The host system can expose this service through a narrow interface---such as a pseudo-device similar to \texttt{/dev/random}, a UNIX-domain socket, or a lightweight request--response API. Deterministic expansion produces far more bits than the physical sampling rate. This allows the seed buffer to serve short bursts of high demand while the RLITBW device continuously refills it; however, expansion increases throughput rather than true entropy, so periodic reseeding and health-test gating remain the security-critical steps.

\subsection*{Security Considerations and Online Health Tests}
\label{sec:health_security}

This section describes a practical deployment architecture for RLITBW as an entropy appliance.
Since the entropy source is macroscopic and optically sensed, a deployment must explicitly consider
both adversarial manipulation and non-adversarial failures (e.g., sensor saturation, pump drift, or
loss of turbulence) that may reduce the entropy rate without immediately breaking basic functionality.

\subsubsection*{Adversary and Failure Model}
We consider a realistic attacker who cannot break the underlying physics of the bubbling process,
but may influence the measurement channel or operating conditions.
Examples include: (i) injecting strong ambient light to saturate the sensor or reduce dynamic range,
(ii) altering optical alignment or occluding the light path,
(iii) mechanically perturbing the air pump (pressure stabilization, forced periodic vibration),
or (iv) inducing environmental changes that reduce turbulence and make trajectories more repeatable.
In addition, benign failures such as a clogged orifice, low water level, bubble ``lock-in'',
or a stuck/saturated sensor can similarly lower entropy.

\subsubsection*{Mitigation Strategy}
\begin{figure*}[!t]
    \centering
    \includegraphics[width=1\linewidth]{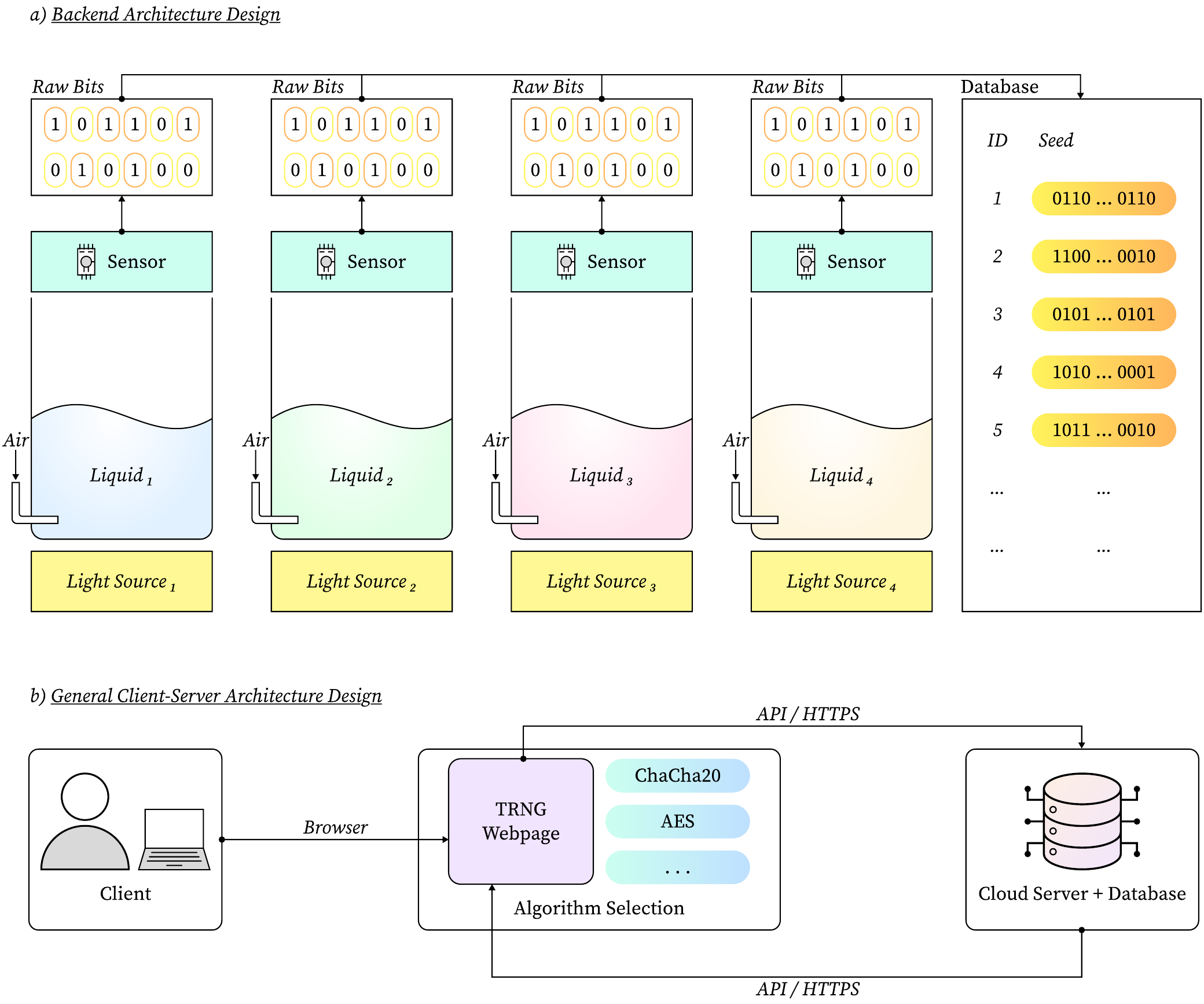}

    \caption{Overall deployment architecture of the proposed system. 
    Multiple RLITBW fluid–optic sensors generate independent raw bitstreams, 
    which are processed by a local backend consisting of a universal-hash 
    extractor (Toeplitz) and a stream or block cipher (ChaCha20/AES). 
    The intermediate high entropy seeds are separately stored in a server-side 
    entropy buffer, and client systems obtain randomness through a 
    lightweight authenticated API or service endpoint.}
    \label{fig:rlitbw-deployment}
\end{figure*}

A practical mitigation strategy is \emph{fail-closed}: the device should stop serving fresh seeds
when health tests indicate a potential entropy collapse.
In the architecture of Fig.~\ref{fig:rlitbw-deployment}, health tests are performed \emph{before} extraction and any on-demand expansion,
so that catastrophic failures are not masked by post-processing.
If a failure is detected, the controller discards current raw blocks, reports an alarm flag to the host,
and pauses seed generation until the physical setup is re-stabilized.

\subsubsection*{Continuous Health Tests (NIST SP 800-90B)}
Modern TRNG guidance emphasizes continuous online tests to detect major failures in entropy sources.
Accordingly, RLITBW deployments can implement the two lightweight health tests recommended for entropy
sources in NIST SP~800-90B: the Repetition Count Test (RCT) and the Adaptive Proportion Test (APT)~\cite{turan2018sp80090b}.
The RCT detects when the digitized samples become ``stuck'' at one value for too long, while the APT detects
when one value becomes much more common than expected within a sliding window~\cite{kelsey90boverview}.

In RLITBW, these tests may be applied to (a) the raw 16-bit sensor samples $X_t$,
or (b) a reduced symbol derived from $X_t$ (e.g., the least-significant byte) to reduce cost while
remaining sensitive to saturation and lock-in.
In addition, simple engineering checks (e.g., saturation-rate monitoring, mean/variance drift alarms)
can be run alongside RCT/APT to quickly detect illumination faults and loss of bubble activity.
Only when all health tests pass is the corresponding block admitted into the extractor input buffer and converted into stored seeds.

\FloatBarrier
\subsubsection*{Discussion}
These protections do not claim immunity against a fully physical attacker with complete control of the environment.
However, explicitly stating an adversary model and integrating fail-closed health testing improves
the robustness of RLITBW as a practical entropy module, and aligns the proposed deployment plan with common TRNG
validation practice~\cite{turan2018sp80090b}.

\textbf{Addressing the throughput}: The architecture naturally scales by running multiple RLITBW units in parallel. Each device contributes its own raw stream, which is individually health-tested and extracted before being merged into a shared entropy pool of stored seeds. Simple strategies such as concatenation (and XOR-based combination as a robustness measure) can be used to combine conditioned outputs. The pool itself may be implemented as a fixed-size RAM ring buffer, with a small software layer enforcing rate limits per client. Adding more devices can increase the entropy refill rate approximately linearly, provided the units are physically independent (e.g., separate vessels/optical paths) and each stream is health-tested and conditioned before pooling, keeping both hardware and software complexity low.

The same design can be extended into a network-facing service. A small server backed by one or more RLITBW modules maintains an in-memory (or database-backed) pool of extracted seeds and exposes an authenticated endpoint that distributes random blocks to external clients by expanding seeds on demand. This model---similar in spirit to existing atmospheric or quantum-based random services---allows virtual machines, containers, or remote applications to share a single physical installation. The overall cost remains low because the RLITBW device itself is inexpensive and easily replaceable, making it a practical entropy appliance for research laboratories, small organizations, or embedded deployments.


\section*{Discussion and Performance Analysis}
\label{sec:discussion_perf}

While the NIST statistical tests confirm the output quality, a practical TRNG must also be evaluated on its throughput, cost-effectiveness, and operational resilience under realistic operating conditions.

\subsection*{Throughput and Latency}
The physical mechanism of the RLITBW system operates on a macroscopic timescale governed by fluid dynamics (bubble formation and rise) and sensor integration time. With a sampling interval of $\Delta t \approx 100$--$200$ ms, the raw physical bit generation rate is on the order of $\sim 80$--$160$ bits per second, depending on the chosen sampling configuration and the effective entropy present in the least-significant bits of the digitized measurements.

This throughput is significantly lower than electronic ring-oscillator TRNGs (often Mbps-scale) or high-speed optical/quantum systems (up to Gbps-scale). However, RLITBW is primarily positioned as a \emph{high-entropy seed generator} rather than a high-throughput bitstream source. In many cryptographic and security applications, physical entropy is required mainly to periodically produce high-quality seeds, after which a deterministic cryptographic primitive can generate application-rate random bits.

In the proposed pipeline, each extracted 256-bit seed is produced from a block of raw samples and can be used to instantiate deterministic expansion (e.g., via ChaCha20) for workloads that demand large volumes of random bits. This design allows the system to meet bursty demand at the interface level while keeping the true-entropy boundary explicit: expansion improves throughput but does not increase the underlying physical entropy beyond what is present in the extracted seeds. If the Practical deployment plan is followed then the throughput problem can be reduced even more by increasing the number of sensors, throughput is increased linearly.

\subsection*{Cost and Complexity Comparison}

A key advantage of the RLITBW system is its extremely low implementation cost achieved using readily available, off-the-shelf components. Table~III presents a qualitative comparison of the proposed system with representative TRNG architectures in terms of cost, implementation complexity, and entropy source.

\begin{table*}[t]
	\centering
	\caption{Comparison of RLITBW with Existing TRNG Technologies}
	\label{tab:comparison}
	\begin{tabular}{|l|c|c|c|}
		\hline
		\textbf{Architecture} & \textbf{Cost} & \textbf{Complexity} & \textbf{Entropy Source} \\
		\hline
		\textbf{QRNG} & Very High & High & Quantum Optics \\
		\hline
		\textbf{FPGA Ring Osc.} & Medium & High & Timing Jitter \\
		\hline
		\textbf{Chaotic Circuits} & Low--Mid & Medium & Nonlinear Voltage \\
		\hline
		\textbf{RLITBW} & \textbf{Very Low} & \textbf{Low} & \textbf{Fluid--Optic Chaos} \\
		\hline
	\end{tabular}
\end{table*}

The complete RLITBW experimental setup was realized using a Raspberry Pi 5 (4 GB), a low-cost BH1750 light intensity sensor, a miniature DC pump powered directly from the Raspberry Pi, and a simple fluid container with tap water. Excluding the laptop used solely for data visualization and post-processing, the total additional hardware cost of the entropy generation system is approximately \textbf{INR 5,000--INR 6,000} (nearly \textbf{USD 60--72}), with the fluid components contributing only a negligible fraction of the total cost.

In contrast, commercial QRNGs typically rely on precision optical assemblies and proprietary hardware, resulting in significantly higher costs, while FPGA-based TRNGs require dedicated development boards and complex design workflows. The minimal hardware requirements, low complexity, and absence of specialized components make RLITBW particularly suitable for low-cost prototyping, educational deployment, and hardware-security applications where accessibility and reproducibility are prioritized over maximum throughput.

\subsection*{Robustness and Failure Modes}
The RLITBW entropy source relies on continuous instability in the bubbling column and time-varying optical refraction. One practical failure mode is \emph{flow regularization} (e.g., partial laminar behavior or repeated bubble trajectories) in which the physical dynamics become more repeatable over time, potentially lowering the entropy rate. Additional failure modes include sensor saturation, optical misalignment, partial occlusion, or significant changes in ambient illumination.

In the present prototype evaluation, the system was operated under conditions intended to preserve turbulent bubble motion and measurable intensity fluctuations. For a deployable TRNG module, robustness should be enforced by (i) ensuring operating settings that maintain sufficient bubble variability, (ii) designing the optics and mechanical mounting to reduce accidental alignment drift, and (iii) explicitly detecting entropy degradation using online health monitoring.

Accordingly, a practical deployment should treat the raw sample stream as the security-critical boundary and apply continuous health tests (e.g., the Repetition Count Test and Adaptive Proportion Test described in NIST SP~800-90B) prior to extraction and any deterministic expansion. On detection of a potential entropy collapse, the correct response is fail-closed behavior: discard affected blocks, stop producing new seeds, and require re-stabilization of the physical source before resuming operation.

Finally, it is important to note that operating in an ``open'' environment can increase natural variability, but it also increases susceptibility to external influence (e.g., strong light injection). For this reason, practical robustness benefits from a balanced approach: allow natural micro-variations while still employing basic physical shielding and continuous monitoring to reduce both accidental failures and intentional manipulation opportunities.

\subsection*{Limitations and Long-Term Stability Considerations}
While the proposed RLITBW system demonstrates statistically robust randomness under the tested conditions, several practical considerations define the scope of the present evaluation. As a macroscopic physical entropy source, the system is influenced by environmental parameters such as ambient temperature, air flow characteristics, mechanical vibration, and optical alignment, each of which can change the statistical properties of the raw signal.

In terms of throughput, the achievable true-entropy rate is constrained by the sampling rate of the optical sensor and the data-acquisition platform. In contrast to high-speed optical chaos systems, the RLITBW architecture prioritizes physical accessibility, simplicity, and interpretability over raw throughput. If higher throughput is required, the most direct strategy is parallelization (multiple independent RLITBW units and/or multiple sensors), combined with strict per-stream health testing and conservative extraction.

Long-term operation does not rely on precise electronic bias points or finely tuned resonance conditions, which suggests practical resilience to moderate parameter drift. Nevertheless, comprehensive long-duration studies (including aging effects, maintenance requirements, and systematic environmental sweeps) remain important future work, together with continuous online health monitoring as recommended by modern TRNG standards (e.g., NIST SP~800-90B).

\section*{Conclusion}

This paper introduced a novel entropy source based on Refracted Light Interaction in Turbulent Bubbling Water (RLITBW). By exploiting the compound chaos of multiphase fluid dynamics and optical refraction, we demonstrated a low-cost, macroscopic system capable of generating high-quality random numbers. A physical–mathematical model was established, attributing the unpredictability to the non-linear divergence of bubble trajectories and the complex caustic networks formed by dynamic liquid lenses.

The raw output, when processed through a Toeplitz-hashing extractor and expanded via a deterministic generator, successfully passed the full NIST SP 800-22 statistical test suite, confirming its statistical indistinguishability from a true random source. A systematic description was also provided of how the RLITBW system can be deployed as a complete randomness-generation framework, including strategies to mitigate its current low-throughput limitation. It was demonstrated that a software-based generator can be developed using locally hosted servers to securely store high-entropy seeds and cipher them when required.

To further reduce predictability, multiple RLITBW setups can be operated in parallel under varying environmental conditions, such as different liquids, bubble injection rates, and lighting parameters. Custom controllers may be used instead of full single-board computers like the Raspberry Pi, and a single sufficiently powerful motor can be employed to induce turbulence in most liquids, thereby lowering system cost.

While the present prototype achieves a throughput of approximately 88 bps, future work will focus on developing a parallelized, multi-sensor array to increase the bit rate into the kilobits-per-second range. A comprehensive analysis of robustness under environmental variations—such as temperature, vibration, liquid properties, light intensity, color, and bubble speed—will also be pursued. Finally, implementing the system as a dedicated entropy source for low-power IoT applications, including password generation and lightweight security services, will provide a practical demonstration of its real-world utility. Ultimately, this work shows that macroscopic chaotic phenomena can be harnessed to build a secure and economically accessible true random number generator.

\bibliography{references}

\section*{Acknowledgements}

The authors would like to express their sincere gratitude to Prof. Kousik Dasgupta for his continuous guidance, valuable suggestions, and constructive discussions throughout the course of this research. His insights and mentorship were instrumental in shaping the direction and quality of this work.

The authors also thank all co-authors for their collaborative efforts, technical contributions, and support during the development and validation of this study.

This research did not receive any specific grant from funding agencies in the public, commercial, or not-for-profit sectors. The project was conducted without external financial support. The minimal experimental and implementation costs involved were supported internally by the Institute, Kalyani Government Engineering College.

\section*{Declaration of Generative AI and AI-assisted Technologies in the Writing Process}

During the preparation of this work, the authors used AI-assisted tools to improve language clarity, grammar, and overall readability. The authors take full responsibility for the content of the publication and confirm that all scientific interpretations, results, and conclusions are their own.

\section*{Availability of Data and Materials}

The datasets generated and analysed during the current study are publicly available in the GitHub repository:
\url{https://github.com/NirjharDebnath/RLITBW-Research-for-finding-Randomness}.

This repository includes raw sensor data, processed bitstreams, and all analysis, optimization, and implementation codes necessary to reproduce the findings of this study.

\section*{Additional information}
Competing interests: The authors declare no competing interests.



\end{document}